\documentclass[fleqn,usenatbib,useAMS]{mnras}

\usepackage{graphicx}	% Including figure files
\usepackage{amsmath}	% Advanced maths commands
\usepackage{amssymb}	% Extra maths symbols
\usepackage{multicol}        % Multi-column entries in tables
\usepackage{bm}		% Bold maths symbols, including upright Greek
\usepackage{pdflscape}	% Landscape pages

\usepackage[T1]{fontenc}
\usepackage{ae,aecompl}

\usepackage{newtxtext,newtxmath}

\usepackage{orcidlink}
\title{Double SSA Spectrum and Magnetic Field Strength of the FSRQ 3C~454.3}

\author[H.-W. Jeong et al.]{
Hyeon-Woo Jeong$^{1, 2}$$^{\orcidlink{0009-0005-7629-8450}}$,
Sang-Sung Lee$^{1, 2}$\thanks{Contact e-mail: \href{mailto:sslee@kasi.re.kr}{sslee@kasi.re.kr}}$^{\orcidlink{0000-0002-6269-594X}}$,
Whee Yeon Cheong$^{1, 2}$$^{\orcidlink{0009-0002-1871-5824}}$,
Jae-Young Kim$^{3, 4}$$^{\orcidlink{0000-0001-8229-7183}}$,
\newauthor
Jee Won Lee$^{2}$$^{\orcidlink{0000-0003-2147-0290}}$,
Sincheol Kang$^{2}$$^{\orcidlink{0000-0002-0112-4836}}$,
Sang-Hyun Kim$^{1, 2}$$^{\orcidlink{0000-0001-7556-8504}}$,
B. Rani$^{5, 2, 6}$$^{\orcidlink{0000-0001-5711-084X}}$,
Jongho Park$^{2}$$^{\orcidlink{0000-0001-6558-9053}}$,
\newauthor
and Mark A. Gurwell$^{7}$$^{\orcidlink{0000-0003-0685-3621}}$
\\
$^{1}$Astronomy and Space Science, University of Science and Technology, 217 Gajeong-ro, Yuseong-gu, Daejeon 34113, Republic of Korea \\
$^{2}$Korea Astronomy and Space Science Institute, 776 Daedeok-daero, Yuseong-gu, Daejeon 34055, Republic of Korea \\
$^{3}$Department of Astronomy and Atmospheric Sciences, Kyungpook National University, Daegu 702-701, Republic of Korea \\
$^{4}$Max-Planck-Institut f\"ur Radioastronomie, Auf dem H\"ugel 69, D-53121 Bonn, Germany \\
$^{5}$NASA Goddard Space Flight Center, Greenbelt, MD 20771, USA \\
$^{6}$Department of Physics, American University, Washington, DC 20016, USA \\
$^{7}$Center for Astrophysics | Harvard $\&$ Smithsonian, 60 Garden Street, Cambridge, MA 02138 USA \\
}

\pubyear{2023}

\begin{document}
\label{firstpage}
\pagerange{\pageref{firstpage}--\pageref{lastpage}}
\maketitle

\begin{abstract}
We present the results of a radio multi-frequency ($\rm 3-340~GHz$) study of the blazar 3C~454.3. After subtracting the quiescent spectrum corresponding to optically thin emission, we found two individual synchrotron self-absorption (SSA) features in the wide-band spectrum. The one SSA had a relatively low turnover frequency ($\nu_{\rm m}$) in the range of $\rm 3-37~GHz$ (lower $\nu_{\rm m}$ SSA spectrum, LSS), and the other one had a relatively high $\nu_{\rm m}$ of $\rm 55-124~GHz$ (higher $\nu_{\rm m}$ SSA spectrum, HSS). Using the SSA parameters, we estimated magnetic field strengths at the surface where optical depth $\tau=1$. The estimated magnetic field strengths were $\rm >7~mG$ and $\rm >0.2~mG$ for the LSS and HSS, respectively. The LSS emitting region was magnetically dominated before the June 2014 $\gamma$-ray flare. The quasi-stationary component (C), $\sim 0.6~{\rm mas}$ apart from the 43 GHz radio core, became brighter than the core with decreasing observing frequency, and we found that component C was related to the LSS. A decrease in jet width was found near component C. As a moving component, K14 approached component C, and the flux density of the component was enhanced while the angular size decreased. The high intrinsic brightness temperature in the fluid frame was obtained as $T_{\rm B, int} \approx (7.0\pm1.0) \times 10^{11}~{\rm K}$ from the jet component after the 2015 August $\gamma$-ray flare, suggesting that component C is a high-energy emitting region. The observed local minimum of jet width and re-brightening behavior suggest a possible recollimation shock in component C. \end{abstract}


%%%%%%%%%%%%%%%%% BODY OF PAPER %%%%%%%%%%%%%%%%%%
% The MNRAS class isn't designed to include a table of contents, but for this document one is useful.
% I therefore have to do some kludging to make it work without masses of blank space.
\begingroup
\let\clearpage\relax
\endgroup
\newpage

\section{Introduction} \label{sec1.intro}
Studying the magnetic field (B-field) properties in the relativistic jets of Active Galactic Nuclei (AGNs) is important for understanding jet formation (and dynamics) and particle acceleration processes. Relativistic jets are launched, collimated, accelerated, and powered by the central engine, in which magnetic fields are considered one of the most critical elements \citep{park2022_review}. Changes in the B-field morphology also influence parsec-scale jet activity, i.e., variability, component ejections, changes in the jet direction, etc., \citep[e.g.][]{lister2009a, lister2013, lister2016, lister2018, lister2021}. Radio and $\gamma$-ray emission, especially in AGN having their jets pointed at us, is produced by relativistic radiating particles (electrons and/or protons) interacting with the jet's magnetic field \citep{rybicki1979}. Leptonic, hadronic, and lepto-hadronic are all possible scenarios for high-energy emission \citep{rani2013, bottcher2007, bottcher2013, rybicki1986, dermer2009}. In this paper, we present a detailed analysis of radio flux and spectral variability of a blazar, 3C454.3, to understand the B-field properties of the emission region(s).

Synchrotron emission encompasses a wide wavelength range, spanning from radio to optical/UV (up to X-rays) wavelengths, and is described by a power-law, as $S_{\nu}=\nu^{\alpha}$. The resulting spectral index, denoted as $\alpha$, falls within the range of $-0.26$ to $-1.73$ for optically thin emission \citep{kangsc2021, kimsh2022}. In scenarios where the density of synchrotron electrons increases, such as in shock-induced jets \citep[e.g.,][]{marscher&gear1985}, the low-energy synchrotron photons are absorbed by the synchrotron electrons, leading to optically thick conditions at lower frequencies. This phenomenon is known as synchrotron self-absorption (SSA). Theoretically, the optically thick spectral index within an SSA region is expected to be $\alpha=2.5$ \citep{rybicki1979}, accompanied by a spectral break at a turnover frequency denoted as $\nu_{\rm m}$. Previous studies have identified the turnover frequency $\nu_{\rm m}$ in a range of approximately $10-170~\rm GHz$ \citep{leejw2017, leejw2020, algaba2018, kangsc2021, kimsh2022}. Examining the spectral properties of the SSA region enables us to investigate the magnetic field strengths within that particular region.

There are two methods to estimate magnetic field strength in relativistic jets, utilizing the opacity effect, including SSA. Firstly, the opacity effect causes the apparent position of a radio core (i.e., a surface of the emission region where $\tau_{\nu}=1$) to vary at different observing frequencies ($\nu$). This means that the radio core moves upstream within the jet as the core-shift effect \citep{lobanov1998}. By measuring the core-shift, it is possible to derive the magnetic field strength in the inner region of the jet (e.g., 1 pc from the jet base, $B_{1}$, and radio core, $B_{\rm core}$) by assuming the equipartition condition, which assumes an equal energy density between particles and the magnetic field. Several studies employing Very Long Baseline Interferometry (VLBI) at multiple wavelengths have estimated the magnetic field strength in relativistic jets of AGNs using the radio core-shift effect \citep[e.g., ][]{osullivan&gabuzda2009, pushkarev2012, fromm2013b, kutkin2014}. These studies have reported that $B_{1}$ falls within the range of approximately $\rm 0.09-2.12~G$.

Alternatively, magnetic field strength can be estimated by utilizing the parameters of synchrotron self-absorption (SSA), assuming a uniform, spherical plasma blob emitting synchrotron radiation \citep{marscher1983}. This method involves measuring the size of the emission region and the flux density at turnover frequency ($\nu_{\rm m}$), which corresponds to the optically thick surface of the emission region. The flux density at $\nu_{m}$ can be obtained from radio multi-wavelength data. Therefore, radio multi-wavelength data is essential for measuring magnetic field strength using the SSA effect. Several studies have employed this method to estimate magnetic field strength \citep[e.g.,][and Cheong et al. 2023 (in prep.)]{leejw2017, algaba2018, leejw2020, kangsc2021, kimsh2022}. Based on their findings, the magnetic field strength of the optically thick emission region typically ranges from a few milli-Gauss (mG) or lower.

The source 3C~454.3 hosts a supermassive black hole (SMBH) with a mass of $3.4\times10^{9}~{\rm M}_{\odot}$ \citep{titarchuk2020}. Classified as a flat-spectrum radio quasar (FSRQ), it exhibits variability across all wavelengths \citep[e.g., ][]{jorstad2010, wehrle2012, jorstad2013, liodakis2020, amaya_almazan2021}. Studies employing VLBI data have revealed that the millimeter core of the source is associated with high-energy emission, particularly during flaring periods. Few studies have delved into the source's spectral properties, such as SSA, at radio and millimeter wavelengths. \citet{kutkin2014} employed quasi-simultaneous observations with the Very Long Baseline Array (VLBA) across a range of frequencies ($4.6 - 43~{\rm GHz}$) and found $B_{\rm 1}=0.4\pm0.2~{\rm G}$ and $B_{\rm core,\,43GHz}=0.07\pm0.04~{\rm G}$. By measuring the time-lag of radio light curves, \citet{mohan2015} reported a magnetic field strength of $B_{1}$=$0.48 \pm 0.21~{\rm G}$ at the jet base. Additionally, \citet{pushkarev2012} estimated $B_{1}$ and $B_{\rm core,15GHz}$ for various AGNs, including 3C~454.3, and found field strengths of 1.13 G and 0.06 G, respectively. While core-shift measurements provide insights into the magnetic field strengths of $B_{1}$ and $B_{rm core}$, the sparsity of quasi-simultaneous multi-frequency VLBI observations limits the ability to measure variation in magnetic field strength. Moreover, \citet{chamani2022} demonstrated a significant time-variable core-shift effect in 3C~454.3, with some epochs showing deviation from the equipartition condition ($k_{r}\colon0.5-1.5$, where $k_{r}$ represents the core-shift index). They estimated the magnetic field strength assuming $k_{r}=1$ (see Figure 10 in their work). In earlier work by \citet{kellermann1969_radiosources}, the superposition of multiple SSA components in the spectra of 3C~454.3 was observed.

High-resolution VLBI observations have provided valuable insights into the structure of 3C~454.3. These observations have revealed the existence of a quasi-stationary component located at a distance of $\rm 0.45 - 0.70 ~mas$ away from the core \citep{pauliny-toth1987, kemball1996, gomez1999, jorstad2005, jorstad2010, jorstad2013, jorstad2017, weaver2022}. Notably, studies by \citet{kemball1996} and \citet{gomez1999} demonstrated that the stationary component exhibits comparable or even higher linear polarization intensity and degree of polarization (DP) compared to the core component, with factors ranging from 3 to 10. Additionally, a consistently aligned electric vector position angle (EVPA) parallel to the jet direction has been reported for over a decade. A possible scenario proposed by \citet{amaya_almazan2021} suggests that the quasi-stationary component might be associated with $\gamma$-ray flares through collisions with jet components.

In this study, we explore the spectral properties, specifically SSA, of the source 3C~454 using radio and millimeter-wavelength data obtained from single-dish and interferometer observations. Our analysis focuses on a unique characteristic of the quasi-stationary component in 3C~454.3. The observations data and analysis methods are described in Section \ref{sec2:Obs&Anal}, while the descriptions on data analyses and the results derived from the analyses are presented in Section \ref{sec3:Data Analysis} and \ref{sec4:Result}, respectively. In Section \ref{sec5:Discussion}, we provide scientific discussions and interpretations of the obtained results. Finally, we summarized our findings and interpretations in Section \ref{sec6:Summary}. Throughout this paper, we adopt the cosmological parameters $\Omega_M = 0.27$, $\Omega_{\Lambda} = 0.73$, $H_0 = 71~{\rm km\,{s}^{-1}\,Mpc^{-1}}$ \citep{komatsu2009}. The redshift of the source is known to be $z=0.859$ \citep{jackson1991}, resulting in a luminosity distance of the source is $D_{\rm L} = 5489~{\rm Mpc}$.

\section{Multi-wavelength Data} \label{sec2:Obs&Anal}
We collected single-dish and array data to establish wide band (2.6 -- 343 GHz) spectra, which were obtained contemporaneously. The mean cadence (the gap between consecutive epochs) of the data sets ranged from 4 days to 31 days. To perform a contemporaneous spectral analysis, we binned the data at each frequency with a bin size of 30 days (i.e., the largest cadence). For example, to establish a spectrum for the epoch 2014 January, all measurements in the epoch and at each frequency were averaged to obtain a weighted mean. These data allowed us to analyze characteristics of the relativistic jet of the source, for example, SSA in radio frequency. Table \ref{table:1:TelandFreq} summarizes the data used in this work.

\subsection{F-GAMMA Data} \label{sec2.1:F-GAMMA}
The Fermi-GST AGN Multi-frequency Monitoring Alliance \citep[F-GAMMA,][]{fuhrmann2016} is a monthly monitoring program for $\sim$60 selected Fermi-GST AGNs using several radio telescopes (Effelsberg 100~m in Germany, IRAM 30~m in Spain and APEX-12~m in Chile). 3C~454.3 had been observed in the program using the Effelsberg 100~m radio telescope in the frequency range of 2.64 -- 345 GHz. We used the F-GAMMA data obtained at 2.64 -- 43 GHz, published by \citet{angelakis2019}. We removed a few outliers that showed a significant flux loss compared to the other data at the same frequency (i.e., KVN single-dish and VLBA-BU-BLAZAR at 43 GHz). The typical fractional uncertainty of the F-GAMMA data was $\sim3.5~\%$ for the whole frequency range.

\subsection{OVRO Data} \label{sec2.2:OVRO}
3C~454.3 has been monitored by the 40 m radio telescope at the Owens Valley Radio Observatory \citep[OVRO\footnote{\url{https://sites.astro.caltech.edu/ovroblazars/}},][]{richards2011} at 15 GHz. The monitoring program started in late 2007, and the mean cadence of the flux measurements for 3C~454.3 is roughly 4 days. We used this data from January 2012 to December 2015 (MJD 55932 -- 57383). The measurement errors in the data are in a range of 0.05 -- 0.86 Jy in this period.

\subsection{CARMA Data} \label{sec2.3:CARMA}
The Combined Array for Research in Millimeter-wave Astronomy (CARMA) was an array of 23 radio telescopes in California. The array can be operated at three atmospheric bands, 1 cm (27 -- 35 GHz), 3 mm (85 -- 116 GHz), and 1 mm (215 -- 270 GHz). The source 3C~454.3 was monitored with a mean cadence of $\sim6$ day at a center frequency of 94.75 GHz in the program called Monitoring of $\gamma$-ray AGN with Radio, Millimeter, and Optical Telescopes \citep[MARMOT\footnote{\url{https://sites.astro.caltech.edu/marmot/}},][]{ramakrishnan2016}. This was a key science program performed in the CARMA. The whole period of the MARMOT program was MJD 56169 -- 57092. The measurement uncertainties are in a range of $\rm 0.02-1.31~Jy$.

\subsection{SMA Data} \label{sec2.4:SMA}
The Submillimeter Array (SMA) is an array of 8-element radio telescopes with baseline lengths from 8~m up to 509~m located in Maunakea in Hawaii. 3C~454.3 is included as a calibrator source at 1.3~mm (225 GHz) and $\rm 850~\mu m$ (343 GHz) in an ongoing monitoring program at the SMA to determine the ﬂuxes of compact extragalactic radio sources, which can be used as calibrators at mm wavelengths \citep{gurwell2007}. Observations of available potential calibrators are from time to time observed for 3 to 5 minutes, with the measured source signal strength calibrated against known standards, typically solar system objects (Titan, Uranus, Neptune, or Callisto).  Data from this program are updated regularly and are available at the SMA website\footnote{\url{http://sma1.sma.hawaii.edu/callist/callist.html}}. The mean cadence of the observations on 3C~454.3 were $\sim7$ day at 1.3 mm and $\sim30$ day at $\rm 850~\mu m$, respectively. We used the SMA data in the period of MJD 55935 -- 57384. The fractional errors, which are driven primarily by telescope systematics (e.g., pointing, system temperature calibration over the full bandpass) and especially by the overall uncertainty in the mm/sub-millimeter flux scale, were $\sim5.5~\%$ and $\sim5.7~\%$ at 1.3 mm and $\rm 850~\mu m$, respectively.

\subsection{ALMA Data} \label{sec2.5:ALMA}
The Atacama Large Millimeter/submillimeter Array (ALMA) consists of 66 antennas operating at 84 -- 950 GHz in the Atacama desert, Chile. The ALMA observes quasars as a calibrator, and the calibrator database\footnote{\url{https://almascience.eso.org/sc/}} is publicly available. Since 3C~454.3 is one of the calibrators for the ALMA observations, we obtained the multi-wavelength data at 91, 103, 233, 337, and 343 GHz. This work used the ALMA data from MJD 56010 to MJD 57385. The mean cadences were 17, 17, 130, 26, and 32 days, respectively. The typical fractional measurement error in the frequency range was $\sim2.5~\%$.

\subsection{KVN Observations and Data Reduction} \label{sec2.6:KVN_SD}
The Korean VLBI Network (KVN) is composed of three identical 21~m-antennas, KVN Yonsei (KYS), KVN Ulsan (KUS), and KVN Tamna (KTN). The KVN has the unique capability of simultaneous multi-frequency observations at four frequencies, 22, 43, 86, and 129 GHz, with a bandwidth of 512 MHz.

Flux density measurements were conducted using cross-scan mode. This measurement method is convenient for measuring pointing offset. The KVN single-dish monitoring observations of blazars are part of the KVN key science program, called MOGABA \citep[MOnitoring of GAmma-ray Bright AGNs,][]{mogaba_lee2013}.
    
We used the GILDAS-CLASS\footnote{\url{https://www.iram.fr/IRAMFR/GILDAS/}} software to calibrate and calculate the source's flux density. Since the observed profile has a background emission from the sky, we removed the baseline flux and fitted it with a Gaussian model. We can correct pointing offsets and calculate peak antenna temperature by fitting the profile. Correction and calculation were done using the equation,
    \begin{equation} {T_{\rm c}} = T_{\rm p}\,e^{4{\rm ln}2\left({\frac {Offset}{Width}}\right)^2}\quad[K], \label{eq:1:Tc} \end{equation}
where $T_{\rm c}$ is the offset--corrected antenna temperature, $T_{\rm p}$ is the peak antenna temperature of the brightness profile, $Offset$ is the antenna pointing offset, $Width$ is the profile width. Then $T_{\rm c}$ was converted into flux density ($S$) as follows$\colon$
    \begin{equation} S = 8\,k_{\rm B} \left(\frac 1{\pi D^{2}A_{\rm e}} \right) T_{\rm c}, \label{eq:2:TtoS} \end{equation}
where $k_{\rm B}$ is the Boltzmann constant, $D=$ 21-m, the diameter of the KVN antenna, $A_{\rm e}$ is the antenna aperture efficiency.

This work uses the KVN single-dish data obtained at 22 and 43 GHz during MJD 55928 -- 57284. Fractional uncertainties were $\sim3~\%$ and $\sim5~\%$ at 22 GHz and 43 GHz, respectively. The mean cadences were $\sim$22 day at 22 GHz and $\sim$19 day at 43 GHz.

    \renewcommand{\arraystretch}{1.2} \begin{table} \caption{Observatories and frequencies of single-dish and array.} \label{table:1:TelandFreq} \begin{tabular*}{\columnwidth}{@{}l@{\hspace*{50pt}}l@{\hspace*{50pt}}l@{}}
    \hline
    Telescope & Frequency (GHz) \\ [2pt]
    \hline
    Effelsberg & $2.64,~4.85,~8.35,~10.45,~14.6,~23.05,~32,~43$  \\ [2pt]
    OVRO       & $15$                                            \\ [2pt]
    KVN        & $22,~43$                                        \\ [2pt]
    CARMA      & $94.75$                                         \\ [2pt]
    SMA        & $225.0,~343.04$                                 \\ [2pt]
    ALMA       & $91.5,~103.5,~233,~337,~343.5$                  \\ [2pt]
    \hline \end{tabular*} \end{table}

\subsection{EVN Data} \label{sec2.7:EVN}
The European VLBI Network\footnote{\url{https://www.evlbi.org/}} (EVN) is composed of 22 radio telescopes located mainly in Europe and Asia. The network provides high angular resolution (e.g., $4\times1.5~{\rm mas^2}$ at 5 GHz) and high sensitivity. The available frequencies of the EVN are in the range of 300 MHz -- 43 GHz. The observed raw and pipeline-calibrated data can be accessed through the public EVN archive\footnote{\url{http://archive.jive.nl/scripts/avo/fitsfinder.php}}. In this work, we used the pipeline-calibrated\footnote{\url{https://www.jive.eu/jivewiki/doku.php?id=parseltongue:grimoire&\#the_evn_pipeline}} data of 3C~454.3 observed at $\rm \sim5~GHz$ to estimate the size of the low-frequency radio core. The additional data descriptions are listed in Table~\ref{table:2:VLBIinfo}.

The data were handled in the \emph{DIFMAP} \citep{shepherd1997} software package for imaging. In the imaging, we used uniform weight rather than natural weight to focus on the core region. The CLEAN image was obtained after repetitive CLEAN and self-calibration processes. Using the calibrated data, we model-fitted the data with circular Gaussian models to obtain the angular size and flux density of the model through the \textit{modelfit} task. We set the initial Gaussian model and fit the model with a few iterations. After that, if a significant intensity remained in the residual map, we set the next Gaussian model and repeated the processes until the peak residual intensity was lower than three times the r.m.s. noise level. The initial models were then fitted until no reduced-$\chi^2$ value change appeared. We dropped a model during the fitting if the model showed a negative flux density or artificially small angular size.

\subsection{VLBA Data} \label{sec2.8:VLBA}
3C~454.3 has also been observed by the Very Long Baseline Array\footnote{\url{https://science.nrao.edu/facilities/vlba}} (VLBA) in the two monitoring programs, Monitoring of Jets in the Active Galactic Nuclei with VLBA Experiments \citep[MOJAVE\footnote{\url{https://www.cv.nrao.edu/MOJAVE/}},][]{lister2021} and VLBA-BU-BLAZAR \citep[BU\footnote{\url{https://www.bu.edu/blazars/VLBAproject.html}},][]{jorstad2005, jorstad2017, weaver2022} at 15 GHz and 43 GHz, respectively. The VLBA data is provided on the website in the form of UVFITS.

At 15 GHz, the source is resolved into extended jet components and the radio core. At 43 GHz, we can focus on the inner region of the radio core via its high angular resolution. Most of the VLBA 43 GHz images show a quasi-stationary component (C) which is about $\rm 0.45-0.7~mas$ away from the core \citep{jorstad2005, jorstad2017, weaver2022}.

We also used additional VLBA data. One was observed in October 2008 at frequencies from 4.6 GHz to 43.2 GHz quasi-simultaneously (BK150). We employed this data to estimate the jet geometry of the source by utilizing core sizes at each frequency (see Section \ref{sec:A:Jet_Geo}). The other observations were conducted in May 2013 at 22 and 43 GHz (BR188), January 2016 (BJ083B), and June 2016 (BJ083C) at 22 GHz. These data were used to investigate the radio spectra of component C (see Section \ref{sec4.2:SSA Locations}).

All the additional VLBA data can be accessed from the National Radio Astronomy Observatory (NRAO) data archive\footnote{\url{https://science.nrao.edu/facilities/vlba/facilities/vlba/data-archive/index}}. We reduced the data with the \emph{AIPS}\footnote{\url{http://www.aips.nrao.edu/index.shtml}} of the NRAO. After loading the correlated raw data into \emph{AIPS}, we first corrected Earth Orientation Parameters (EOPs) by using the \emph{VLBAEOPS} task. Dispersive delays caused by the ionosphere were corrected by using the ionospheric models obtained from NASA Crustal Dynamics Data Information System\footnote{\url{https://cddis.nasa.gov/}} (CDDIS) via the procedure of \emph{VLBATECR}. The sampler voltage offsets in the cross-correlation spectra were corrected using the auto-correlation spectra. Instrumental phase offsets and delays across the IFs were corrected via a scan of 3C~454.3. The remaining residual delays and rates were found and calibrated via \emph{FRING} in \emph{AIPS} (i.e., global fringe fitting). Amplitude calibration was performed using the observed gain curves and system temperatures. The task \emph{BPASS} was used with a scan of a bright source  to calibrate the bandpass shapes. We used the same scan used for the instrumental phase offset calibration. The final data was imaged with \emph{DIFMAP,} just as we followed for the EVN data imaging.

    \renewcommand{\arraystretch}{1.2} \begin{table} \caption{The description of the additional VLBI observations.} \label{table:2:VLBIinfo} \fontsize{7.9}{7.9} \begin{tabular}{clccc}
    \hline
    Date       & Exp. Code            & Beam Size                & $\rm \nu_{obs}$ \\ [0pt]
    {}         & (Instru.)            & $({\rm mas \times mas})$ & $({\rm GHz})$   \\ [0pt]
    \hline
    2013-05-23 & $\rm EG062C\,(EVN) $ & $0.89 \times 4.93$       & $4.99 $         \\ [0pt]
    2013-05-27 & $\rm EG062D\,(EVN) $ & $1.25 \times 4.80$       & $4.99 $         \\ [0pt]
    2014-03-03 & $\rm EP088D\,(EVN) $ & $1.00 \times 1.07$       & $4.99 $         \\ [0pt]
    2014-03-05 & $\rm EG062E\,(EVN) $ & $0.87 \times 5.07$       & $4.99 $         \\ [0pt]
    2013-05-02 & $\rm BR188\,(VLBA) $ & $0.37 \times 0.82$       & $22.24$         \\ [0pt]
    2014-06-10 & $\rm EL048B\,(EVN) $ & $1.05 \times 1.20$       & $4.99 $         \\ [0pt]
    2014-10-26 & $\rm EG084B\,(EVN) $ & $0.97 \times 4.46$       & $4.99 $         \\ [0pt]
    2015-03-16 & $\rm ES076\,(EVN)  $ & $0.65 \times 3.83$       & $6.67 $         \\ [0pt]
    2015-06-06 & $\rm EK035\,(EVN)  $ & $0.95 \times 1.09$       & $4.99 $         \\ [0pt]
    2015-11-01 & $\rm ER038\,(EVN)  $ & $0.95 \times 3.32$       & $4.99 $         \\ [0pt]
    2016-01-28 & $\rm BJ083B\,(VLBA)$ & $0.24 \times 0.64$       & $24.58$         \\ [0pt]
    2016-02-26 & $\rm ES079A\,(EVN) $ & $2.73 \times 4.75$       & $6.67 $         \\ [0pt]
    2016-06-05 & $\rm ES079B\,(EVN) $ & $0.69 \times 3.20$       & $6.67 $         \\ [0pt]
    2016-06-06 & $\rm BJ083C\,(VLBA)$ & $0.26 \times 0.63$       & $23.98$         \\ [0pt]
    \hline \end{tabular} \end{table}

\subsection{GMVA Data} \label{sec2.9:GMVA}
The global millimeter VLBI array\footnote{\url{https://www3.mpifr-bonn.mpg.de/div/vlbi/globalmm/}} (GMVA) is an array consisting of sensitive radio telescopes (8 VLBA stations, GBT, Effelsberg, Pico Veleta, Onsala, Metsaehovi, Yebes, and KVN) at 86 GHz, and offers very high angular resolution. This work used the calibrated GMVA data observed in September 2015. The data reduction procedures for the GMVA data are described in \citep[][see also references therein]{casadio2019}. The GMVA observation was conducted in dual-polarization mode, although we used only the total intensity data because the reason for using the GMVA data was to investigate the jet structure of 3C~454.3.

%%%%%%%%%%%%%%%%%%%% Data Analysis %%%%%%%%%%%%%%%%%%%%
\section{Data Analysis} \label{sec3:Data Analysis}

\subsection{Quiescent Flux Density Determination} \label{sec3.1:Quiescent Flux}
\begin{figure*} \includegraphics[width=2\columnwidth]{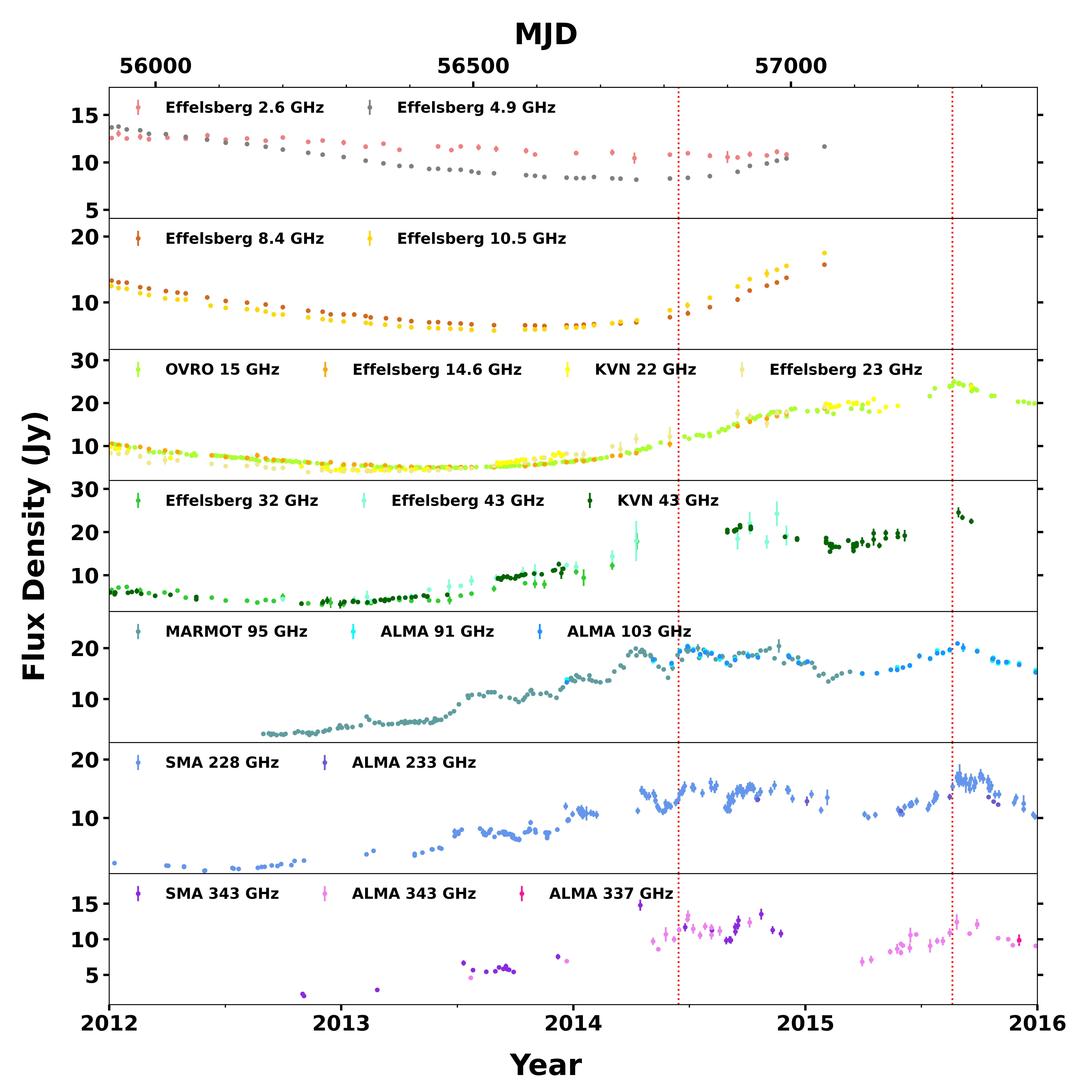}
    \caption{Multi-wavelength light curves were obtained from 2.6 to 343 GHz. The observing frequencies and the telescopes are provided as legends in each panel. The two vertical red-dotted lines indicate two $\gamma$-ray flares, which peaked on $\sim$MJD 56823 \citep{gamma2014}, and $\sim$MJD 57254 \citep{gamma2015}, respectively.} \label{fig:1:MW_SDandA} \end{figure*}

For spectral analysis of the source, we used multi-frequency radio data in the range of 2.6 -- 343 GHz from 2012 to 2016, as shown in Figure \ref{fig:1:MW_SDandA}. The source showed a significant flux variation ranging from $\rm 1.3-25~Jy$ in all wavelengths. The minimum flux density of the source at 225 GHz was 1.3 Jy in 2012, which is the lowest flux density observed with the SMA over a decade.

To accurately analyze the spectra of the variable components in the light curve (i.e., to subtract the quiescent flux), we used an exponential flare model to decompose the light curve. The decomposition of a light curve is described by \citet{valtaoja1999} and applied as follows:
    \begin{equation}
    S(t) = \begin{cases} S_{\rm max}\,e^{(t-t_{\rm max})/t_{\rm r}} , &     t < t_{\rm max} \\ 
    S_{\rm max}\,e^{(t_{\rm max}-t)/1.3t_{\rm r}} , & t > t_{\rm max}, \end{cases}  \label{eq:3:Flare_model} \end{equation} 
where $S_{\rm max}$ is the peak flux density of a flare, $t_{\rm max}$ is the time when a flare peaks, $t_{\rm r}$ is the rising timescale of a flare. We used a 1.3  ratio of decaying to rising timescales by following \citet{valtaoja1999} and found good fitting results with the fixed ratio.

    \begin{figure} \includegraphics[width=\columnwidth]{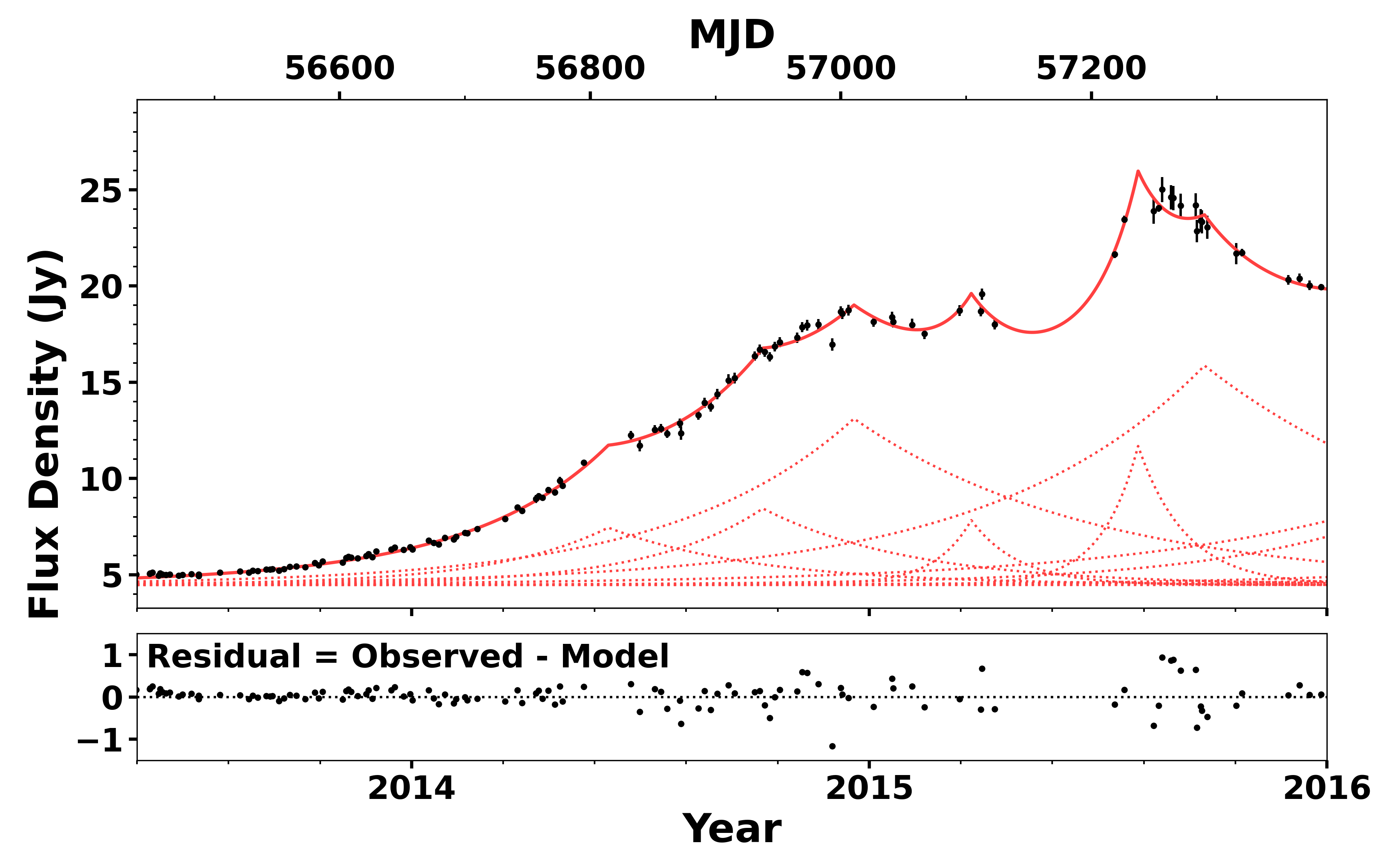}
    \caption{15 GHz OVRO flare decomposition. \emph{Upper}$\colon$The red solid line and dotted lines indicate the sum of all the fitted flares and individual flares, respectively. The minimum BIC value is obtained as 516 by 13 flare models. The period is $\rm MJD~56448 - 57388$. The estimated quiescent flux is $\rm 4.47\pm0.07~Jy$, denoted as the black horizontal line. \emph{Lower}$\colon$Flux difference between observed measurements and flare decomposition. The black dotted horizontal line is where the residual flux density is 0.} \label{fig:2:Flare_Decomposition} \end{figure}

We fitted the first flare to the earliest peak in the light curve and then investigated the remaining flux density after subtracting the fitted flare model from the light curve. For the remaining light curve, we fitted the next subsequent flare and repeated this process until no flare-like peak  appeared from the residual light curve (i.e., a peak residual flux density $\leq 3 \times \sigma_{\rm rms}$, where $\sigma_{\rm rms}$ is the r.m.s. noise level of a light curve). The flare models were fitted using the \emph{emcee} python library \citep{emcee2013}, which is an implementation of the Markov Chain Monte Carlo (MCMC). After that, we removed one flare that had an artificially short $t_{\rm r}$ or low $S_{\rm max}$ at one time and then estimated both reduced-$\chi^2$ and Bayesian information criterion (BIC) value to select the most statistically reasonable total number of flares. BIC is a method of selecting the best-fit model from the data by adding a penalty when the number of model parameters increases. Reduced-$\chi^2$ decreases when the number of flares increases. Meanwhile, the BIC value showed a minimum value in the best-fit result. As a representative example, we show the flare decomposition of the 15 GHz light curve in Figure \ref{fig:2:Flare_Decomposition}.

\begin{figure} \includegraphics[width=\columnwidth]{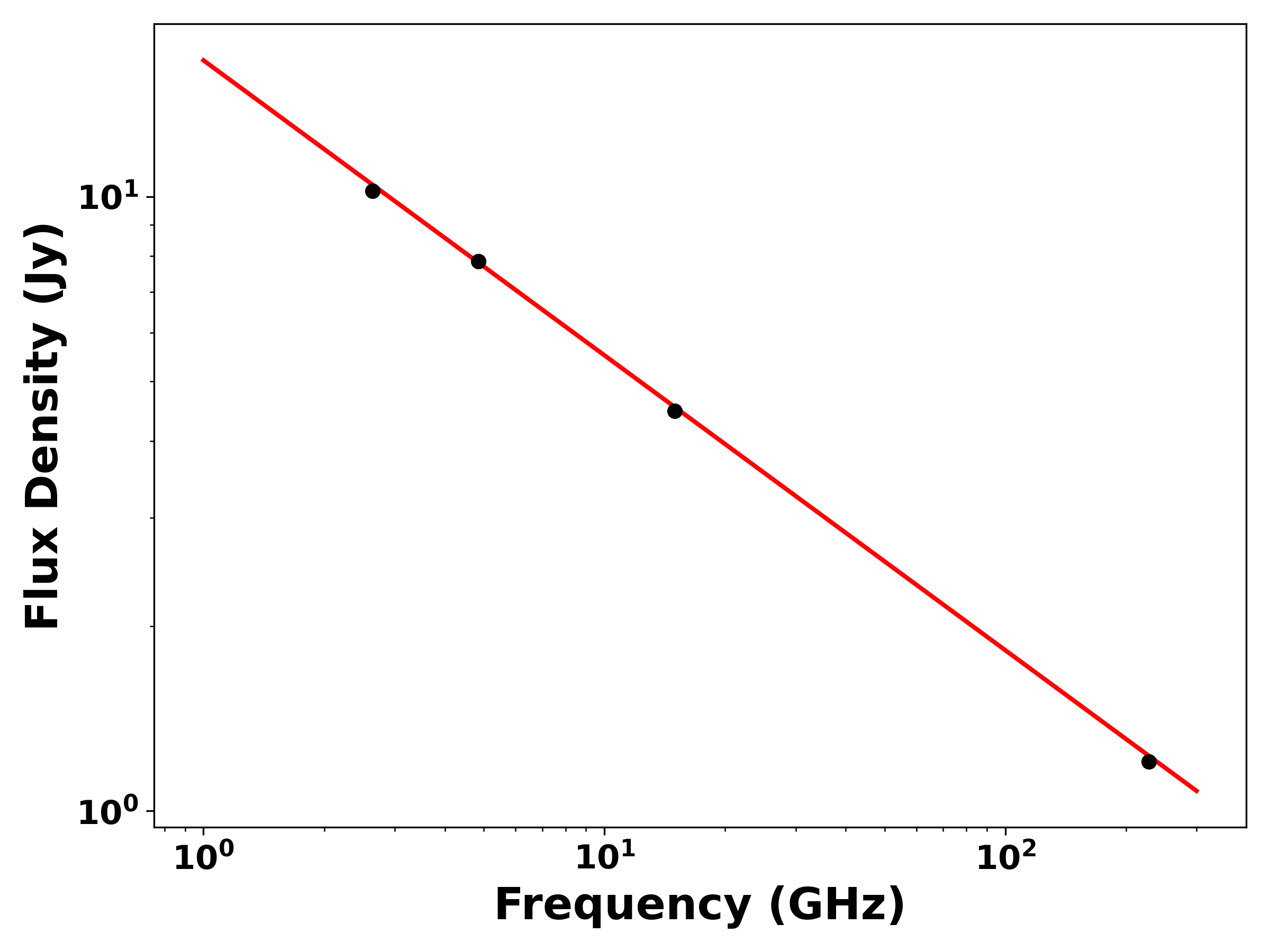}
    \caption{Quiescent spectrum fitting result using radio data at four frequencies (i.e., 3, 5, 15, and 225 GHz). Black dots indicate estimated quiescent fluxes from flare decomposition. The solid red line indicates the estimated quiescent spectrum of 3C~454.3.} \label{fig:3:QuiescentFlux} \end{figure}

We selected light curves at 15 and 225 GHz to estimate the quiescent spectrum of the source because, at these frequencies, observation cadences are better than at other frequencies, and the quiescent period of the source is covered. Lower frequency (3 and 5 GHz) light curves were also used to cover the low-frequency range of the quiescent spectrum. The estimated quiescent fluxes were $\rm 10.2\pm0.11~Jy$ (3 GHz), $\rm 7.85\pm0.06~Jy$ (5 GHz), $\rm 4.47\pm0.07~Jy$ (15 GHz) and $\rm 1.2\pm0.02~Jy$ (225 GHz). The quiescent spectrum of the source was fitted to a simple power-law model ($F(\nu)=C_{\rm q}\nu^{\alpha}$). The model parameters resulted in $C_{\rm q}=16.66\pm0.07~{\rm Jy}$ and $\alpha=-0.48$ with very small uncertainty. The estimated quiescent fluxes and the quiescent spectrum model are shown in Figure \ref{fig:3:QuiescentFlux}. A broken power-law model diverges when used to fit the observed minimum flux densities.

\subsection{SSA Spectra Fitting} \label{sec3.2:SSA Spectra}
The light curves had different cadences from each other (3 days -- a month). Light curves were binned with a 30 days interval, the largest gap, to investigate their spectral variation. After binning, the quiescent spectrum was subtracted from the light curves. At $\rm \sim340~GHz$, we used the flux density extrapolated from the obtained quiescent spectrum. From the quiescent-subtracted spectra, we found two spectral peaks in 2013$\colon$$\nu_{\rm m}$ of the  lower turnover frequency SSA spectrum (LSS) was between 3 -- 5 GHz, and of the higher turnover frequency SSA spectrum (HSS) was between 55 -- 124 GHz. In 2014, on the other hand, the LSS seen in 2013 disappeared (i.e., a single spectral peak), and the high-frequency part of the HSS flattened, which may imply a superposition of two or multiple SSA components in high-frequency spectra (see below for further discussion). 

In this work, we fitted the variable spectra with two SSA models rather than three or more models due to the lack of data points. The SSA spectrum model was introduced by \citet{turler1999} as follows:
    \begin{equation}
    S(\nu)=S_{\rm m} \left(\frac {\nu}{\nu_{\rm m}}\right)^{\alpha_{\rm thick}} 
    {\frac {1-e^{-\tau_{\rm m}(\nu/\nu_{\rm m})^{\alpha_{\rm thin}-\alpha_{\rm thick}}}}{1-e^{-\tau_{\rm m}}}}, \label{eq:4:SSA_model} \end{equation} 
where $S_{\rm m}$ and $\tau_{\rm m}$=$\frac{3}{2}\times\left(\sqrt{1-\frac{8\alpha_{\rm thin}}{3\alpha_{\rm thick}}}-1\right)$ are the flux density of the SSA spectrum and the optical depth at $\nu_{\rm m}$, respectively. $\alpha_{\rm thin}$ and $\alpha_{\rm thick}$ are the spectral indices of the optically thin and thick parts. In this work, we fixed $\alpha_{\rm thick}$=$2.5$ because, theoretically, the optically thick intensity is proportional to the source function, $S_{\nu} = {P(\nu)} / {4 \pi \kappa_{\nu}} \propto \nu^{5/2}$ \citep{rybicki1979}. $\alpha_{\rm thin}$ was limited from -3 to 0, and $\nu_{\rm m}$ was constrained from $\rm 0.1-10~GHz$ and $\rm 10-200~GHz$ for the LSS and the HSS, respectively. We note that the frequency boundaries were applied to only the first epoch. For the consecutive epochs, $\nu_{\rm m}$ of the previous epoch was used as an initial parameter with a broader boundary limit (e.g., $\rm 0.1-50~GHz$ for the LSS).

\begin{figure} \includegraphics[width=\columnwidth]{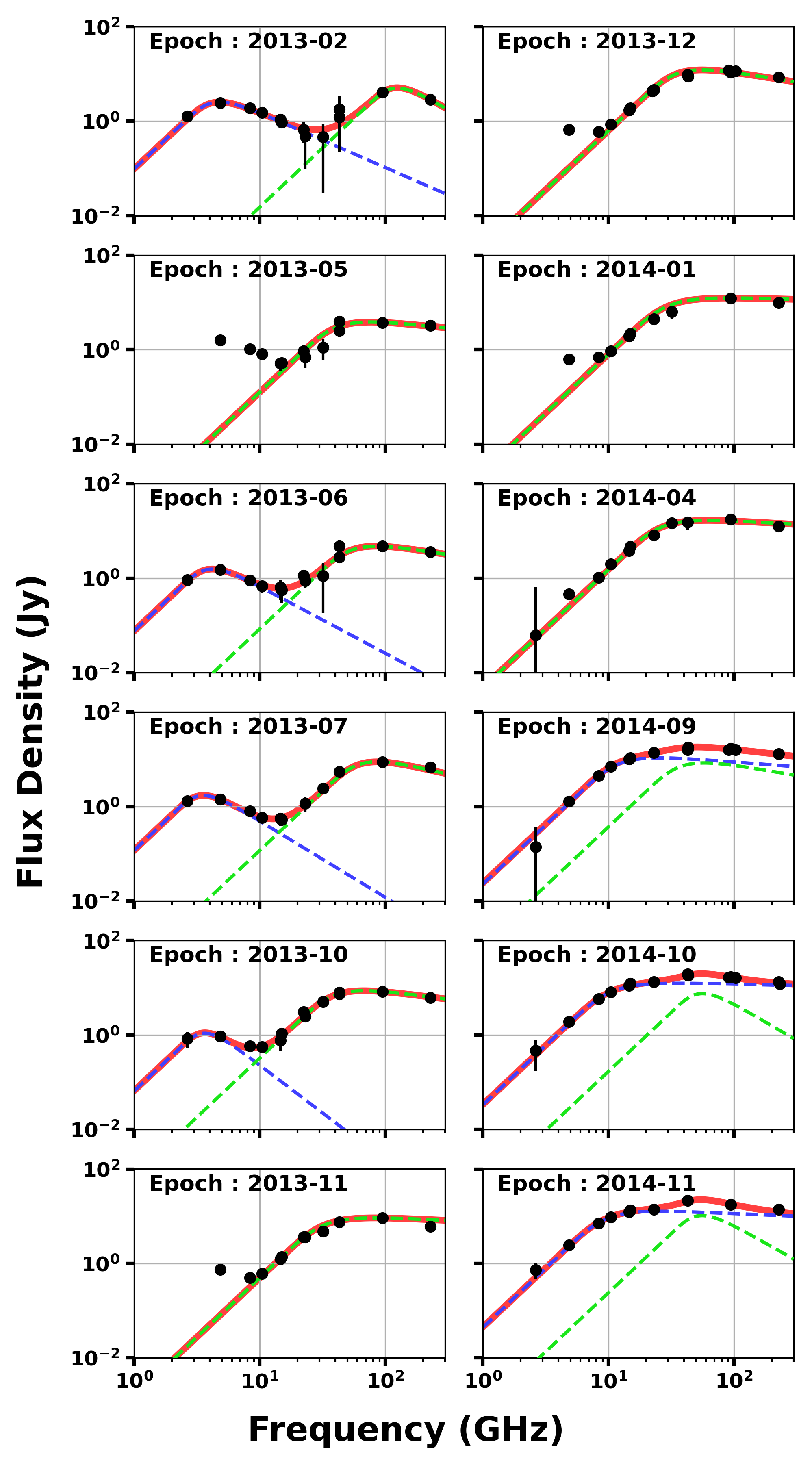}
    \caption{Source spectra with double SSA components from February 2013 to November 2014. Green and blue dashed lines indicate the LSS and the HSS, respectively. The solid red line indicates the sum of the two SSA spectra. Black dots indicate quiescent-emission-subtracted fluxes. Dates are denoted in each panel.} \label{fig:4:SSA_Spectra} \end{figure}

Peak flux density $S_{\rm m}$ was set to vary in a positive range. Although the spectra in late 2014 ($2014-09$, $2014-10$, $2014-11$) were likely a single SSA component, a single SSA model failed to fit the observed spectra properly due to the flattening in high frequency. For example, especially in $2014-11$, the flux density at 43 GHz was higher than the  $S_{\rm m}$ of the LSS model, possibly indicating  another SSA emission. For this reason, we chose a double SSA model to fit the observed spectra. Note that for the epochs when 3 GHz data was not available ($2013-05$, $2013-11$, $2013-12$, and $2014-01$), only the single SSA model (i.e., HSS) was used because the $\nu_{\rm m}$ of the LSS could not be determined appropriately. Additionally, when the flux density at 3 GHz was comparable to the uncertainty of the estimated quiescent flux density at 3 GHz ($2014-04$), only the HSS was modeled for further analysis. The best-fit results are shown in Figure \ref{fig:4:SSA_Spectra}. The best-fit parameters are summarized in Table \ref{table:3:SSA Parameters}.

    \renewcommand{\arraystretch}{1.2} \begin{table*} \caption{Best-fit SSA model parameters. $S_{\rm m}$ is the flux density at a turnover frequency $\nu_{\rm m}$. $\alpha_{\rm thin}$ is the spectral index for an optically thin part.} \label{table:3:SSA Parameters} \fontsize{7.2}{7.2} \begin{tabular}{ccccccccccc}
    \hline
    Epoch           & {}                      & {}                      & LSS                     & {}                      & {}                      & {}                      & {}                         & HSS                     & {}                       & {}                       \\
    {}              & ${S_{\rm m}}$           & $\nu_{\rm m}$           & $\alpha_{\rm thin}$     & $\theta_{\rm FWHM}$     & ${T_{\rm B, obs}}^{*}$  & ${S_{\rm m}}$           & $\nu_{\rm m}$              & $\alpha_{\rm thin}$     & $\theta_{\rm FWHM}$      & ${T_{\rm B, obs}}^{*}$   \\
    (Year -- Month) & (Jy)                    & (GHz)                   & {}                      & ($\rm \mu as$)          & ($\rm 10^{11}~K$)       & (Jy)                    & (GHz)                      & {}                      & ($\rm \mu as$)           & ($\rm 10^{11}~K$)        \\
    \hline
    $2013-02$       & $2.54 ^{+0.15}_{-0.16}$ & $4.76 ^{+0.37}_{-0.31}$ & $-1.16^{+0.14}_{-0.17}$ & $917\pm88$              & $1.63\pm0.36$           & $5.04 ^{+1.58}_{-0.87}$ & $123.37^{+7.35} _{-12.89}$ & $-1.59^{+0.70}_{-0.97}$ & $23\pm2$                 & $7.98\pm2.36$            \\ [3pt]
    $2013-05$       &          {$-$}          &          {$-$}          &          {$-$}          & {-}                     & {-}                     & $3.84 ^{+0.20}_{-0.17}$ & $76.64 ^{+17.71}_{-8.77}$  & $-0.28^{+0.13}_{-0.16}$ & $34\pm2$                 & $6.92\pm2.06$            \\ [3pt]
    $2013-06$       & $1.55 ^{+0.12}_{-0.12}$ & $4.13 ^{+0.39}_{-0.30}$ & $-1.41^{+0.31}_{-0.38}$ & $1194\pm125$            & $0.78\pm0.19$           & $4.72 ^{+0.19}_{-0.19}$ & $85.38 ^{+7.19} _{-6.09}$  & $-0.42^{+0.10}_{-0.10}$ & $15\pm1$                 & $34.04\pm5.85$           \\ [3pt]
    $2013-07$       & $1.71 ^{+0.17}_{-0.16}$ & $3.51 ^{+0.19}_{-0.19}$ & $-1.61^{+0.28}_{-0.31}$ & $1625\pm203$            & $0.64\pm0.17$           & $8.83 ^{+0.16}_{-0.16}$ & $85.79 ^{+6.19} _{-4.88}$  & $-0.61^{+0.05}_{-0.05}$ & $30\pm2$                 & $16.19\pm2.75$           \\ [3pt]
    $2013-10$       & $1.09 ^{+0.23}_{-0.18}$ & $3.57 ^{+0.62}_{-0.43}$ & $-2.02^{+0.49}_{-0.51}$ & $1573\pm192$            & $0.42\pm0.15$           & $8.65 ^{+0.45}_{-0.34}$ & $67.40 ^{+9.03} _{-5.50}$  & $-0.35^{+0.13}_{-0.16}$ & $40\pm2$                 & $14.85\pm2.75$           \\ [3pt]
    $2013-11$       &          {$-$}          &          {$-$}          &          {$-$}          & {-}                     & {-}                     & $9.21 ^{+0.15}_{-0.14}$ & $82.82 ^{+20.27}_{-10.50}$ & $-0.13^{+0.06}_{-0.06}$ & $36\pm2$                 & $12.81\pm3.90$           \\ [3pt]
    $2013-12$       &          {$-$}          &          {$-$}          &          {$-$}          & {-}                     & {-}                     & $12.22^{+0.29}_{-0.29}$ & $55.51 ^{+1.07} _{-1.01}$  & $-0.43^{+0.03}_{-0.03}$ & $42\pm4$                 & $27.80\pm4.88$           \\ [3pt]
    $2014-01$       &          {$-$}          &          {$-$}          &          {$-$}          & {-}                     & {-}                     & $12.41^{+0.41}_{-0.33}$ & $93.28 ^{+65.35}_{-27.91}$ & $-0.08^{+0.06}_{-0.13}$ & $58\pm7$                 & $5.21\pm4.19$            \\ [3pt]
    $2014-04$       &          {$-$}          &          {$-$}          &          {$-$}          & {-}                     & {-}                     & $16.70^{+1.17}_{-1.07}$ & $62.10 ^{+9.39} _{-5.31}$  & $-0.16^{+0.06}_{-0.06}$ & $71\pm8$                 & $10.62\pm3.07$           \\ [3pt]
    $2014-09$       & $10.75^{+1.67}_{-2.06}$ & $25.56^{+9.16}_{-7.44}$ & $-0.20^{+0.11}_{-0.29}$ & $96\pm5$                & $21.67\pm10.64$         & $8.39 ^{+3.79}_{-2.06}$ & $58.04 ^{+8.29} _{-6.71}$  & $-0.45^{+0.21}_{-0.58}$ & $78\pm3$                 & $4.95\pm1.61$            \\ [3pt]
    $2014-10$       & $12.43^{+0.54}_{-0.69}$ & $36.45^{+7.49}_{-6.18}$ & $-0.06^{+0.02}_{-0.04}$ & $98\pm3$                & $11.80\pm3.27$          & $7.53 ^{+0.96}_{-0.88}$ & $55.53 ^{+2.53} _{-3.20}$  & $-1.56^{+0.64}_{-0.84}$ & $80\pm3$                 & $4.64\pm0.63$            \\ [3pt]
    $2014-11$       & $12.70^{+0.57}_{-0.72}$ & $25.93^{+3.64}_{-3.92}$ & $-0.11^{+0.03}_{-0.06}$ & $136\pm8$               & $12.51\pm2.97$          & $10.32^{+2.62}_{-2.59}$ & $55.20 ^{+7.69} _{-5.96}$  & $-1.50^{+0.65}_{-1.90}$ & $94\pm4$                 & $4.70\pm1.24$            \\ [0pt]
    \hline \multicolumn{11}{l}{* Note$\colon$The uncertainties in $T_{\rm B, obs}$ were propagated using the mean error values of the upper and lower uncertainties of other parameters.} \end{tabular} \end{table*}

We note that some information could be lost because of binning if variability is shorter than 30 days. Shorter bin sizes may allow a more effective trace for spectral evolution. However, the low mean cadences ($20-30$ days) at low frequencies ($\rm < 15~GHz$) prevented us from applying shorter bin sizes. In addition, shorter bin sizes would not allow us to collect enough data points to perform the double-SSA model fitting.

\subsection{Magnetic Field Strength Estimation} \label{sec3.3:B Field Estimation}
Following \citet{marscher1983}, the magnetic field strength ($B_{\rm SSA}$) of an SSA region is estimated, assuming a uniform sphere, as follows$\colon$
    \begin{equation} \begin{aligned}
    B_{\rm SSA}= 10^{-5} b(\alpha) \left[{\frac {S_{\rm m}} {\rm 1~Jy}}\right]^{-2} \left[{\frac {d_{\rm m}} {\rm 1~mas}}\right]^{4} \left[{\frac {\nu_{\rm m}} {\rm 1 GHz}}\right]^{5} \left({\frac {1+z} {\delta}}\right)\;[\rm G],
    \label{eq:5:B_SSA} \end{aligned} \end{equation} 
where $d_{\rm m}=1.8 \times \theta_{\rm FWHM}$ is the size of the SSA region at $\nu_{\rm m}$. $\theta_{\rm FWHM}$ is obtained using the jet geometry, which can be estimated through multi-wavelength VLBI observations (see Appendix \ref{sec:A:Jet_Geo} for more details) and is multiplied by 1.8 to change the full-width at half maximum (FWHM) of a Gaussian model into the size of a uniform sphere. $\delta$ and $z$ are the Doppler factor of a jet component and the redshift of the source, respectively, and $b(\alpha)$ is a factor dependent on $\alpha_{\rm thin}$. 

We adopted $\delta_{\rm LSS}=20.3\pm1.8$ and $\delta_{\rm HSS}=28.6\pm2.1$ as the Doppler factors, which are the mean values of components $\Delta r>0.4~{\rm mas}$ and $\Delta r\leq0.4~{\rm mas}$, respectively, where $\Delta r$ is the distance of the component from the core \citep{jorstad2005, jorstad2017, weaver2022}. The factor $b(\alpha)$ can vary from 1.8 to 3.8 \citep[see Table 1 in ][]{marscher1983}. We note that Equation \ref{eq:5:B_SSA} differs from the original equation in \citet{marscher1983} \citep[i.e., using the factor $(1+z)/\delta$, as introduced in][]{algaba2018}, because we assume both the radio core and the component C are in a steady state rather than exhibiting a moving feature, except for the last three epochs in the LSS (K14 is a moving knot, see Section~\ref{sec3.4:Features of core and C} and \ref{sec4.2:SSA Locations}).

In addition to $B_{\rm SSA}$, following \citet{kataoka&stawarz2005}, who estimated the magnetic field strength ($B_{\rm EQ}$) under an equipartition condition where the particle energy density and magnetic field energy density were equivalent, we have
    \begin{equation} \begin{aligned}
        B_{\rm EQ}= 0.123 \times  \eta^{{\frac {2} {7}}} \times {\left(1+z\right)}^{\frac {11} {7}} \times \left[{\frac {D_{\rm L}} {\rm 100~Mpc}}\right]^{\frac {-2} {7}} \left[{\frac {\nu_{\rm m}} {\rm 5 GHz}}\right]^{\frac {1} {7}} \\ \times \left[ \frac {S_{\rm m}} {\rm 100~mJy}\right]^{\frac {2} {7}} \left[ \frac {{d_{\rm m}}} {0.3''}\right]^{\frac {-6} {7}}
        {\delta}^{\frac {-5} {7}} \quad[{\rm mG}],
    \label{eq:6:B_EQ} \end{aligned} \end{equation}
where $\eta$ is a jet composition factor (e.g., $\eta=1 \colon {\rm leptonic}$, $\eta=1836 \colon {\rm hadronic}$), $D_{\rm L}$ is the luminosity distance to the source. In this work, we used two pure cases (i.e., leptonic and hadronic) to estimate the field strengths.

    \begin{figure} \includegraphics[width=\columnwidth]{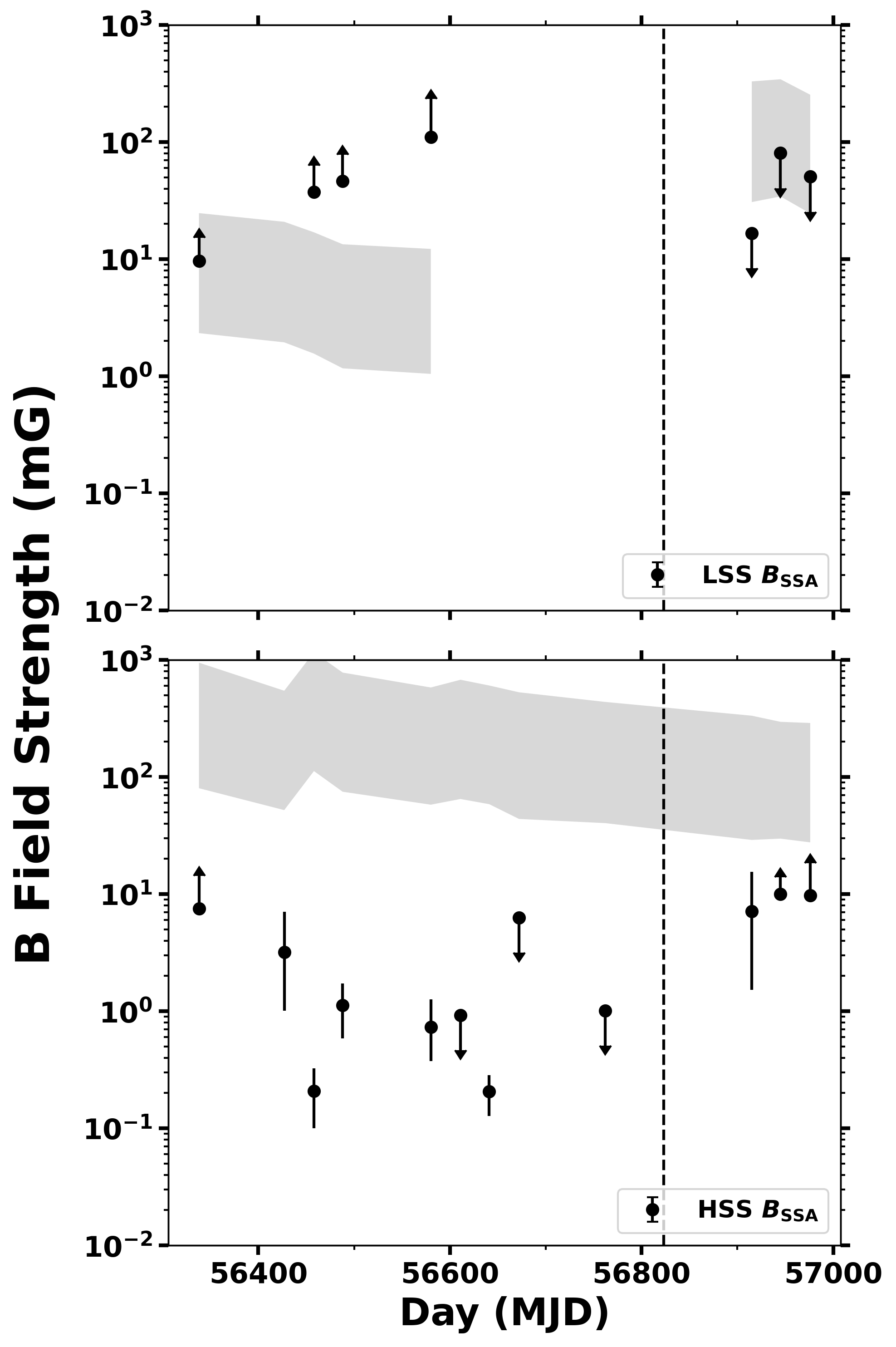}
    \caption{Time variation of B-field strengths for the LSS (\emph{Upper}) and the HSS (\emph{Lower}). Black dots indicate the estimated $B_{\rm SSA}$ in units of mG. The grey shaded areas show the corresponding range of the time-interpolated $B_{\rm EQ}$ for $\rm \eta=1-1836$.} The black dashed vertical line indicates the 2014 June $\gamma$-ray flare. The arrows show upper and lower limits. \label{fig:5:B_Field} \end{figure}

The estimated magnetic field strengths are shown in Figure \ref{fig:5:B_Field}. For the LSS (\emph{Upper} panel) emitting region before the June 2014 $\gamma$-ray flare \citep[peaked $\rm \sim MJD~56823$, hereafter, $\gamma_{2014}$,][]{gamma2014}, the estimated $B_{\rm SSA, LSS}~(\rm > 37~mG)$ was higher than $B_{\rm EQ, LSS}~(\rm 1-25~mG)$ except for the first epoch, February 2013, which depended on $\eta\colon$$B_{\rm SSA, LSS} > B_{\rm EQ, LSS}~(\eta=1)$ and $B_{\rm SSA, LSS} \lesssim B_{\rm EQ, LSS}~(\eta=1836)$. These indicate that the LSS region possibly deviates from an equipartition condition. In other words, the magnetic field energy density dominates the total energy density in the jet. After the $\gamma_{2014}$, $B_{\rm EQ, LSS}~{\rm (25-350~mG)}$ significantly increased. In this period, since the upper limits of the estimated $B_{\rm SSA, LSS}~(<81~{\rm mG})$ are in the range of $B_{\rm EQ, LSS}$, the energy density dominance may depend on $\eta$.

For the HSS emitting region, most of the estimated $B_{\rm SSA, HSS}~{\rm (0.2-7~mG)}$ is lower than $B_{\rm EQ, HSS}~{\rm (27-1200~mG)}$ indicating that the emission region is dominated by particle energy density. All of the estimated $B_{\rm SSA}$ and $B_{\rm EQ}$ are listed in Table~\ref{table:4:B_Strength}.

    \renewcommand{\arraystretch}{1.5}
    \begin{table*} \caption{The estimated B-Field strengths. The upper and lower uncertainties were propagated individually through the standard error propagation model using the SSA parameters.} \label{table:4:B_Strength} \fontsize{7.5}{7.5} \begin{tabular}{ccccccc}
    \hline
    Epoch           & $B_{\rm SSA,LSS}$ & $B_{\rm EQ,LSS}$          & $B_{\rm EQ,LSS}$            & $B_{\rm SSA,HSS}$       & $B_{\rm EQ,HSS}$           & $B_{\rm EQ,HSS}$              \\
    {}              & {}                & {$\eta=1$}                & {$\eta=1836$}               & {}                      & {$\eta=1$}                 & {$\eta=1836$}                 \\
    (Year -- Month) & (mG)              & (mG)                      & (mG)                        & (mG)                    & (mG)                       & (mG)                          \\
    \hline
    $2013-02$       & $>9.69  $         & $2.61^{+0.28}_{-0.28}$    & $22.39^{+2.37}_{-2.37}$     & $>7.50$                 & $95.28^{+15.20}_{-13.47}$  & $815.70^{+130.13}_{-115.29}$  \\ [3pt]
    $2013-05$       & $-$               & $-$                       & $-$                         & $3.19^{+3.89}_{-2.19}$  & $57.87^{+5.77}_{-5.51}$    & $495.42^{+49.42}_{-47.18}$    \\ [3pt]
    $2013-06$       & $>37.39 $         & $1.77^{+0.21}_{-0.21}$    & $15.18^{+1.80}_{-1.79}$     & $0.21^{+0.12}_{-0.11}$  & $123.95^{+11.76}_{-11.74}$ & $1061.16^{+100.69}_{-100.49}$ \\ [3pt]
    $2013-07$       & $>46.38 $         & $1.37^{+0.20}_{-0.20}$    & $11.72^{+1.70}_{-1.70}$     & $1.13^{+0.60}_{-0.54}$  & $82.83^{+8.10}_{-8.09}$    & $709.15^{+69.37}_{-69.22}$    \\ [3pt]
    $2013-10$       & $>110.09$         & $1.24^{+0.19}_{-0.18}$    & $10.61^{+1.63}_{-1.57}$     & $0.73^{+0.53}_{-0.36}$  & $62.88^{+4.94}_{-4.81}$    & $538.31^{+42.30}_{-41.18}$    \\ [3pt]
    $2013-11$       & $-$               & $-$                       & $-$                         & $<0.93$                 & $71.88^{+7.09}_{-6.75}$    & $615.37^{+60.66}_{-57.80}$    \\ [3pt]
    $2013-12$       & $-$               & $-$                       & $-$                         & $0.20^{+0.08}_{-0.08}$  & $64.50^{+6.00}_{-6.00}$    & $552.23^{+51.41}_{-51.41}$    \\ [3pt]
    $2014-01$       & $-$               & $-$                       & $-$                         & $<6.31$                 & $52.75^{+8.97}_{-7.59}$    & $451.64^{+76.82}_{-64.99}$    \\ [3pt]
    $2014-04$       & $-$               & $-$                       & $-$                         & $<1.01$                 & $45.69^{+5.37}_{-5.29}$    & $391.13^{+45.95}_{-45.30}$    \\ [3pt]
    $2014-09$       & $<16.71 $         & $34.64^{+3.88}_{-3.91}$   & $296.57^{+33.25}_{-33.44}$  & $7.10^{+8.31}_{-5.58}$  & $33.96^{+4.99}_{-3.34}$    & $290.74^{+42.71}_{-28.61}$    \\ [3pt]
    $2014-10$       & $<80.46 $         & $37.28^{+2.92}_{-2.88}$   & $319.17^{+25.03}_{-24.67}$  & $>9.97$                 & $32.12^{+2.39}_{-2.35}$    & $274.97^{+20.50}_{-20.14}$    \\ [3pt]
    $2014-11$       & $<50.85 $         & $27.12^{+2.53}_{-2.55}$   & $232.17^{+21.63}_{-21.84}$  & $>9.69$                 & $30.68^{+3.07}_{-3.03}$    & $262.64^{+26.25}_{-25.90}$    \\ [0pt]
    \hline \end{tabular} \end{table*}

\subsection{Milli-arcsecond Scale Structures} \label{sec3.4:Features of core and C}
The milli-arcsecond (mas) source structure was investigated using VLBI observations at 5 and 43 GHz. To obtain the flux density of an emitting region, we model-fitted the EVN 5 GHz and the VLBA 43 GHz data with circular Gaussian models. An elliptical Gaussian model was used when a circular Gaussian model could not describe the long baseline \textit{uv} visibilities (i.e., core region). For example, in the VLBA observations from January 2014 to June 2014, the elliptical Gaussian model describes the observed visibilities better than a circular one. We obtained core flux densities from the fitted models. 

\begin{figure} \includegraphics[width=\columnwidth]{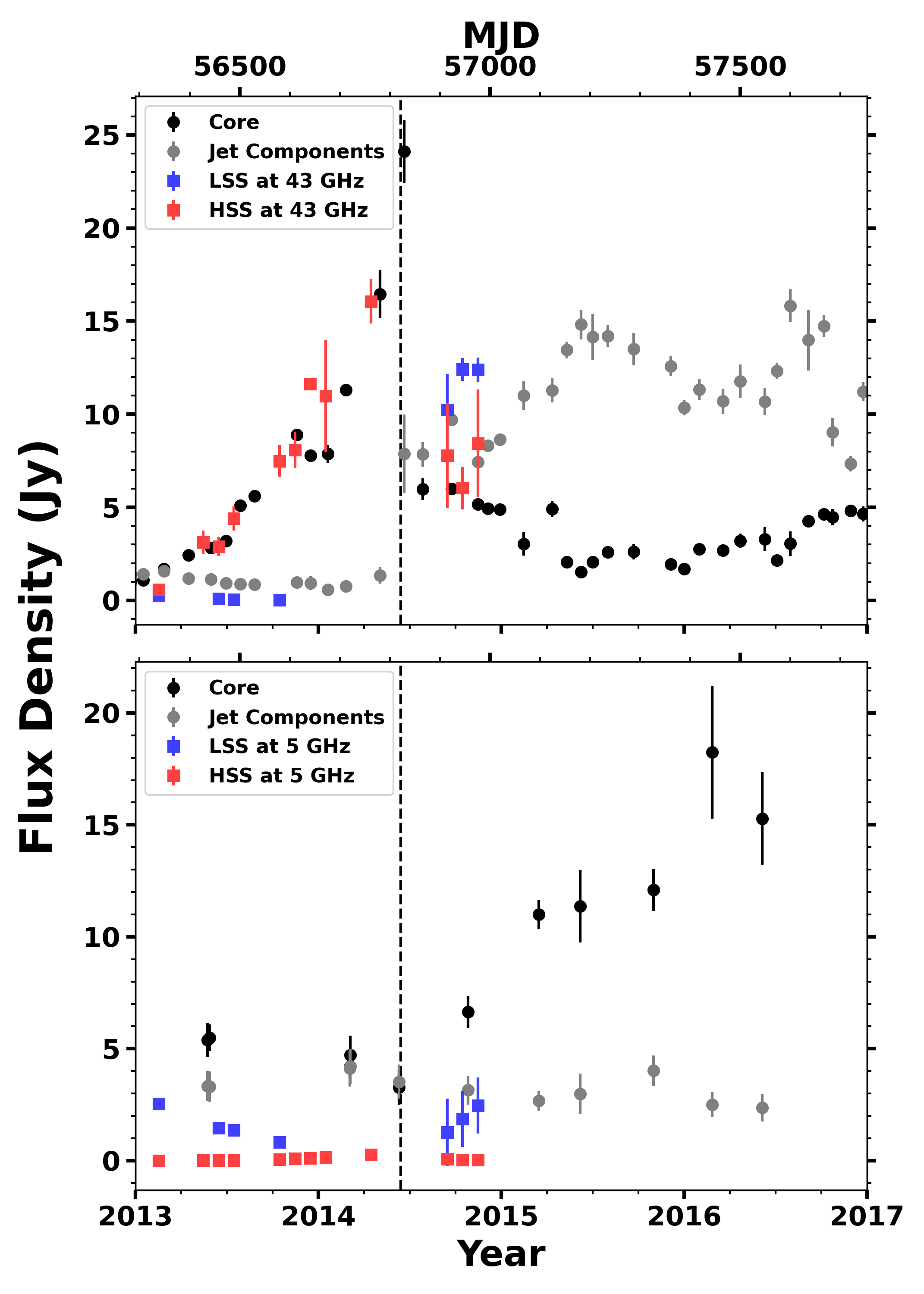}
    \caption{Flux comparison at 43 GHz (\emph{Upper}) and 5 GHz (\emph{Lower}) between SSA-inferred and model-fit measurements. The black dashed vertical line indicates the 2014 June $\gamma$-ray flare. Uncertainties on SSA-inferred flux densities are propagated through the standard error propagation model using the obtained SSA parameters.} \label{fig:6:Core-SSA} \end{figure}

As shown in Figure \ref{fig:6:Core-SSA}, the flux variations of the core and jet components were compared to those of the SSA-inferred flux density (i.e., the inferred flux density from the SSA models at a specified frequency). We note that the grey dots in the Figure represent the sum of the flux density of the extended jet models. The errors in the models were determined using residual r.m.s. noise level following \citet{fomalont1999}. The blue and red dots are the SSA-inferred flux density of the LSS and the HSS, respectively. The uncertainties of the SSA-inferred flux densities were propagated using the standard error propagation model.

\begin{figure*} \includegraphics[width=2\columnwidth]{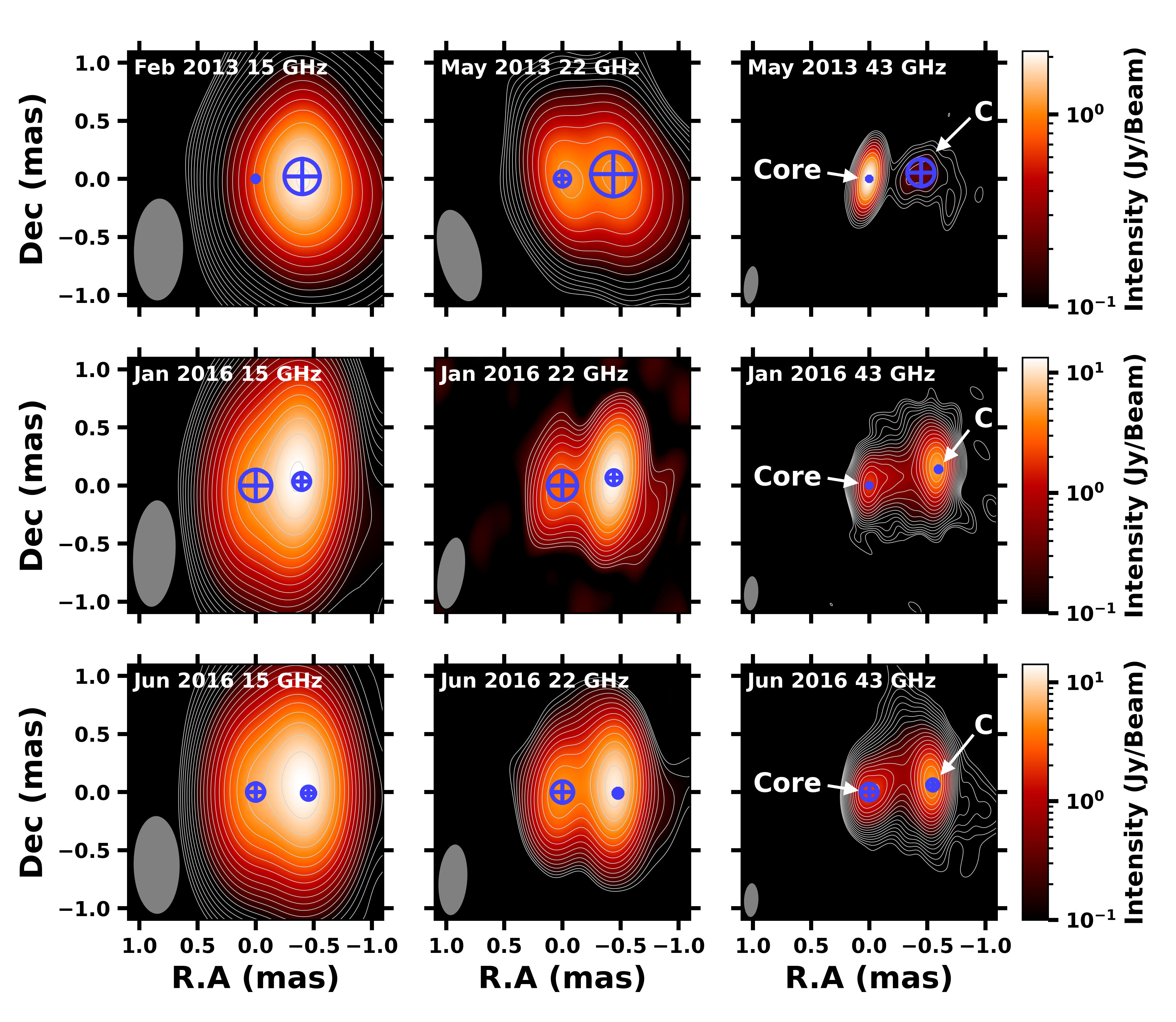}
    \caption{Images of 3C~454.3 with VLBA observations at three frequencies, 15, 22, and 43 GHz. The radio cores are aligned to the (0,0) position. Blue circles indicate Gaussian models for radio core and component C. Observation date and observing band are noted in the upper left side in each panel. color and contours indicate intensity for each map, and the contours start at three times the r.m.s. of the residual map and increase by factors of 1.4. Synthesized beams are attached on the lower left side of each map as a grey ellipse.} \label{fig:7:MW_VLBA} \end{figure*}

\begin{figure} \includegraphics[width=\columnwidth]{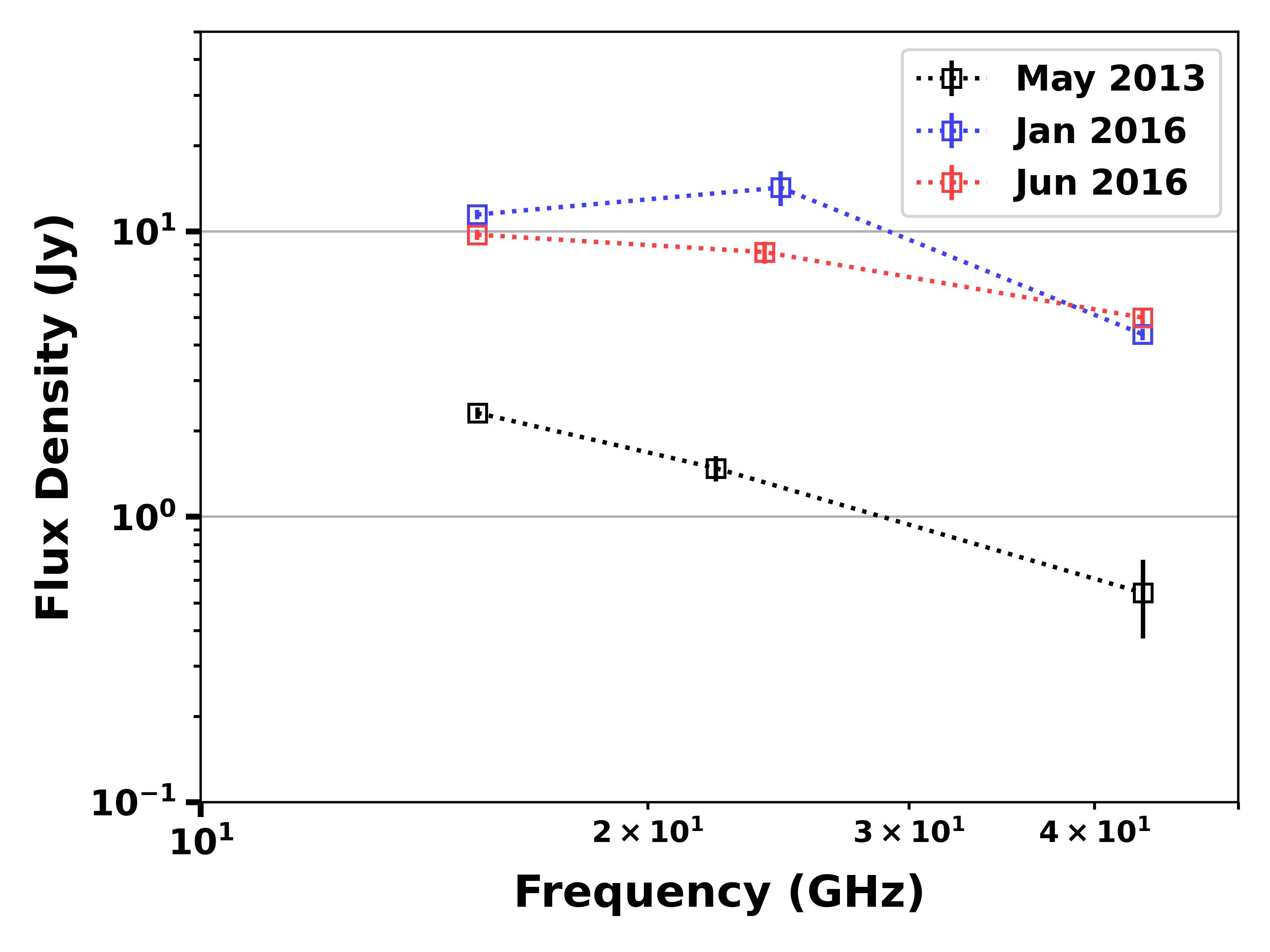}
    \caption{Radio spectra of component C in three epochs. For each epoch, different colors represent the spectrum (see the labels).} \label{fig:8:Spectra_C} \end{figure}

3C~454.3 has been reported to have a highly polarized quasi-stationary component (component C), away from the core by $\rm \sim0.6~mas$, whose flux density is comparable to that of the 43 GHz core or even brighter \citep{kemball1996, jorstad2005, jorstad2013}. To investigate whether the LSS emitting region is related to component C (i.e., to see if component C has a synchrotron self-absorption spectrum), we analyzed spectral features using the VLBA data. In the VLBI spectral analysis, we used the additional three VLBA data at 22 GHz (see Section~\ref{sec2.8:VLBA}). The observations across three frequencies (15, 22, and 43 GHz) are quasi-simultaneous within a few weeks, except for the first epoch at 15 GHz (i.e., in February 2013 at 15 GHz). Component C was successfully identified in all the maps, according to its parameter conditions ($\rm \sim0.6~mas$ of radial distance from the core, $\sim\rm-90^{\circ}$ of position angle, and flux density). The intensity maps with the Gaussian models and the spectra of component C are shown in Figure \ref{fig:7:MW_VLBA} and \ref{fig:8:Spectra_C}, respectively.
    
\citet{liodakis2020} reported that a new jet component, K14 \citep[or B12 in][]{weaver2022}, was ejected from the 43 GHz core. The proper motion of K14 resulted in as $\rm 0.471\pm0.003~mas/yr$ \citep{liodakis2020} and $\rm 0.466\pm0.002~mas/yr$ \citep{weaver2022}. We cross-checked the proper motion of the K14 using the \textit{modelfit} results, yielding $\rm 0.47\pm0.01~mas/yr$. At this proper motion, K14 arrived at $\rm 0.6~mas$ in August 2015 (${\rm \sim MJD}~57238$), when another $\gamma$-ray flare occurred \citep[peaked $\rm \sim MJD~57254$, hereafter, $\gamma_{2015}$,][]{gamma2015}. This indicates that the $\gamma_{2015}$ is possibly associated with an interaction between the components C and K14 as suggested by \citet{amaya_almazan2021}. After that, K14 remained around $\rm \sim0.6~mas$, rather than showing a significant motion, being a quasi-stationary component.

\subsection{Jet Ridgeline Analysis} \label{sec3.5:Ridgeline}
\begin{figure} \includegraphics[width=\columnwidth]{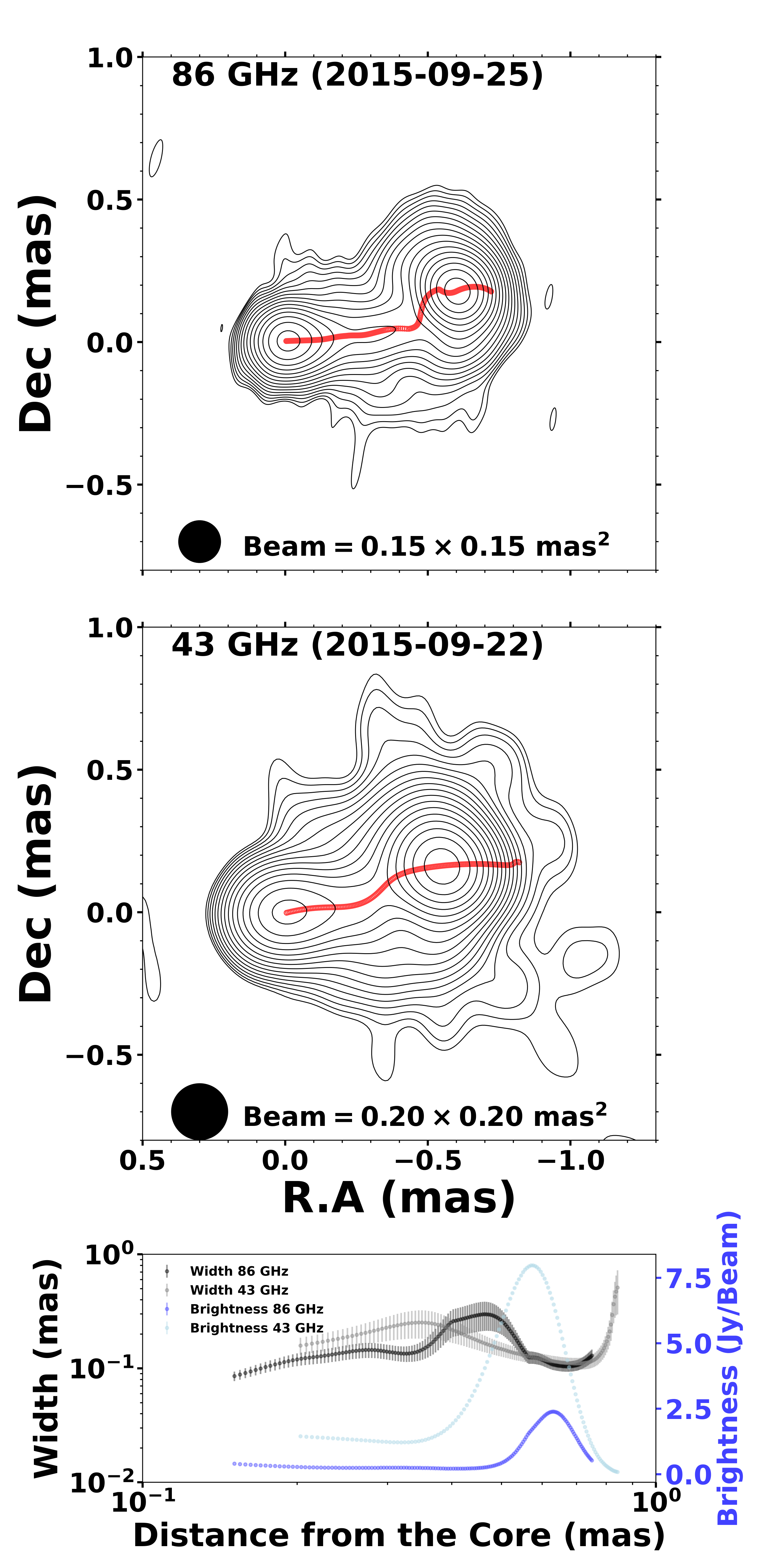}
    \caption{The images show the ridgelines at 86 GHz (\textit{Top}) and at 43 GHz (\textit{Middle}) observed in September 2015. The \textit{Bottom} panel shows the estimated jet width and brightness along the ridgelines at each frequency.} \label{fig:9:Jet_Ridge} \end{figure}

The jet ridgeline, a trail of plasma blobs in the jet, was investigated using the VLBA observations at 43 and 86 GHz, conducted in September 2015,  a month after the $\gamma_{2015}$ flare. The two observations are quasi-simultaneous with a time difference of three days, therefore providing robustness to the analysis by allowing a comparison of the two results. We used radial coordinates to find the jet ridgeline and selected one pixel with the highest brightness in every radial distance interval from the core. The selected pixels form the jet ridgeline, as shown in Figure \ref{fig:9:Jet_Ridge}. After forming the ridgeline, we sliced the maps transversely to the ridgeline. The obtained brightness profile from every slice was fitted to a Gaussian model to estimate jet width and peak brightness. To ensure 5-$\sigma$ detection at half maximum of the profile,  we used a 10-$\sigma$ level as the threshold (i.e., we dropped fit results if the peak brightness was lower than 10-$\sigma$ level, $\sigma_{\rm 43GHz} \approx {\rm 6~mJy}$, and $\sigma_{\rm 86GHz} \approx {\rm 3~mJy}$). 

The convolution effect of the restoring beam can be removed by deconvolving the FWHM of the restoring beam from a fitted Gaussian FWHM $\colon$ $W_{\rm J}=\sqrt{{\theta_{\rm G}}^{2}-{\theta_{\rm B}}^{2}}$, where $W_{\rm J}$ is the jet width, $\theta_{\rm G}$ and $\theta_{\rm B}$ are the full-width at half maximum of the Gaussian model and the restoring beam, respectively. The uncertainties of the jet width along the jet ridgeline were calculated by following \citet{pushkarev2017}. Figure \ref{fig:9:Jet_Ridge} shows the maps with the jet ridgelines denoted by red circles in the \emph{Top} and \emph{Middle} panels, and both the estimated jet width (black dots) and brightness (blue dots) along the jet ridgeline in the \emph{Bottom} panel. Bright colors indicate those at 43 GHz. Note that the \textit{Bottom} panel shows the estimated jet width and brightness along the ridgeline that is further apart than the restoring beam size. The distance of the jet ridgeline was obtained by calculating the position of the pixel.

The jet ridgeline analysis showed a local minimal jet width and a local maximal brightness near component C ($\rm \sim0.6~mas$) at 43 and 86 GHz (Figure~\ref{fig:9:Jet_Ridge}). The de-projected distance of those sites can be calculated by adopting the viewing angle $\theta_{\rm v}\approx1.3^{\circ}$ \citep{jorstad2005}, and the mass of the SMBH $M_{\rm SMBH}\approx3.4\times10^{9}~{\rm M_{\odot}}$ \citep{titarchuk2020}. The calculated de-projected distance was $\sim6\times10^{5}~r_{\rm g}$. However, note that the de-projected distance may not be accurate since both the core-shift effect and the viewing angle vary with time.

\subsection{Stacking the 43 VLBI Maps} \label{sec3.6:Stacking VLBI Maps}
In order to investigate the time-averaged polarimetric characteristics of the jet, we stacked the 38 VLBA 43 GHz data observed between 2013 and 2016 and recorded in dual-polarization mode. Before stacking the maps, all were aligned by setting the core position to (0,0). Aligned maps were then convolved with a common circular restoring beam, whose size was $\rm 0.2 \times 0.2~mas^{2}$.

\begin{figure} \includegraphics[width=\columnwidth]{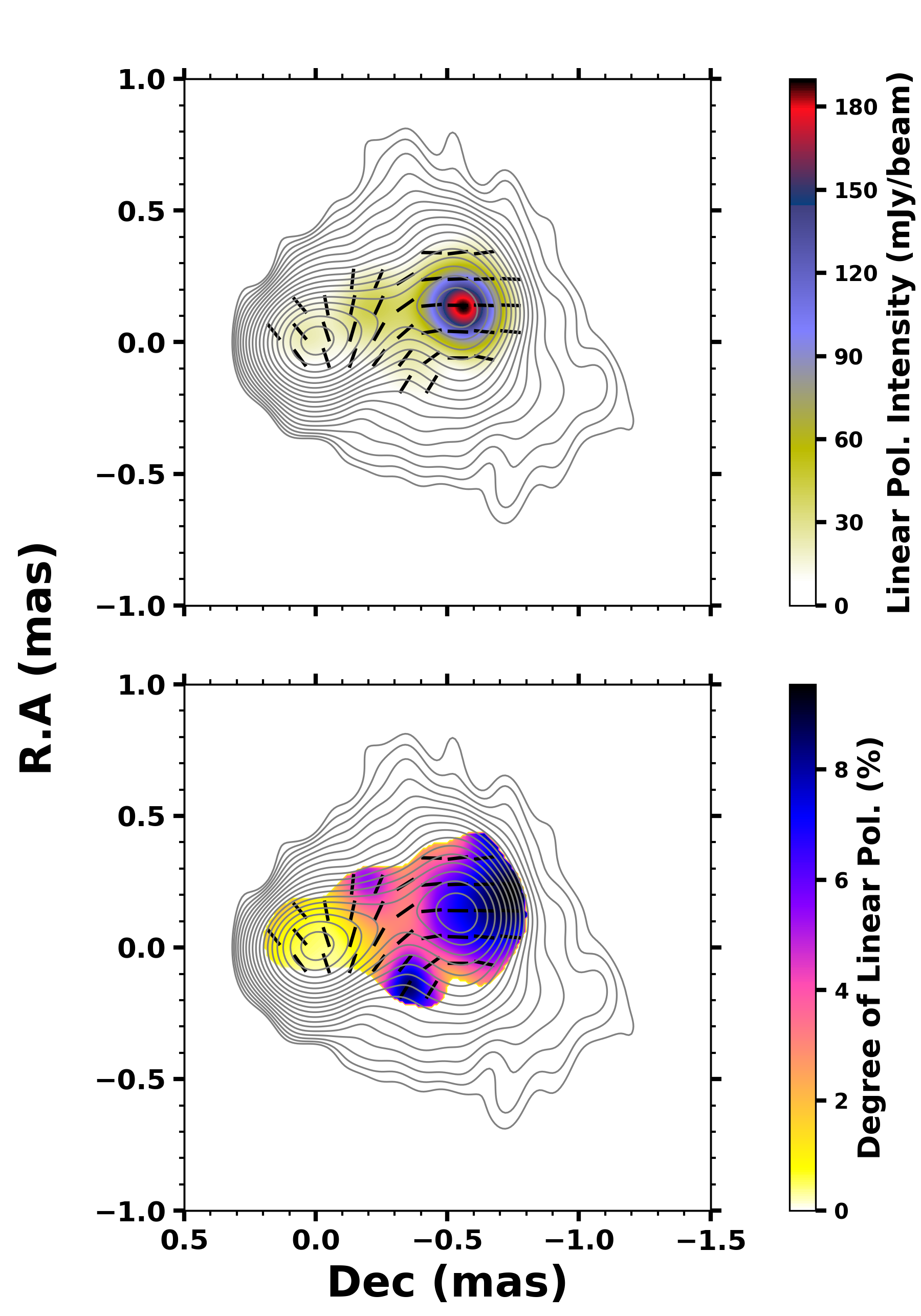}
    \caption{Stacked linear polarization maps for 3C~454.3. Contours indicate Stokes \textit{I} intensity. The black bars indicate polarization angles in the area. The period is the same as Figure \ref{fig:9:Jet_Ridge}. Color maps indicate polarization intensity (\emph{Upper}) and degree of polarization (\emph{Lower}).} \label{fig:10:Stacked_Pol_Map} \end{figure}

To generate the polarization maps, we stacked the maps using Stokes parameters, \textit{I}, \textit{Q}, and \textit{U}. Then the stacked maps were used to compute the maps of polarization intensity and DP. The stacked maps are presented in Figure~\ref{fig:10:Stacked_Pol_Map}.

Based on the core-shift effect studies on 3C~454.3 using about five years of data \citep{kutkin2014, chamani2022}, we found the mean distance offset of cores at 22 and 43 GHz to be $\rm \sim0.05~mas$. Although we assumed the core-shift on the 43 GHz core was negligible based on the small offset between 22 and 43 GHz, the map alignments may not be accurate as the averaged core-shift offsets were not zero, and the core-shift effect varies with time.

\subsection{Brightness Temperature Calculation}
\label{sec3.7:Brightness Temperature}
In the VLBI observations, the brightness temperature in the observer's frame ($T_{\rm B, obs}$) can be higher than the equipartition brightness temperature \citep[$T_{\rm EQ} \approx 5 \times 10^{10}~{\rm K}$,][]{readhead1994} and even the inverse-Compton limit \citep[$T_{\rm IC} \approx 10^{12}~{\rm K}$,][]{kellermann1969_radiosources} due to a strong Doppler boosting effect \citep{lee2008, pushkarev2012Tb, lee2016, jorstad2017, homan2021}. $T_{\rm B, obs}$ of a jet component is calculated using the equation$\colon$
    \begin{equation} \begin{aligned}
    T_{\rm B, obs} = 1.22 \times 10^{12} \frac {S} {{\theta_{\rm FWHM}}^{2}\,\nu^{2}} \quad[{\rm K}],
    \label{eq:7:T_B} \end{aligned} \end{equation}
where $S$ and $\theta_{\rm FWHM}$ are the flux density (Jy) and FWHM (mas) of the component at frequency $\nu$ (GHz).

We calculated the $T_{\rm B, obs}$ values of the two SSA emitting regions (i.e., LSS and HSS), using the best-fit parameters as shown in Table~\ref{table:3:SSA Parameters}.

We also calculated the $T_{\rm B, obs}$ of K14, the most prominent jet component in our data period. The estimated parameters (flux density, distance from the core, position angle, angular size, and the observed brightness temperature) spanning about two years are summarized in Table~\ref{table:5:K14} and shown in Figure~\ref{fig:11:K14}.

    \renewcommand{\arraystretch}{1.2} \begin{table} \caption{The estimated parameters of the moving knot K14.} \label{table:5:K14} \fontsize{5.9}{5.9} \begin{tabular}{cccccc}
    \hline
    Day     & S           & R              & $\theta_{\rm FWHM}$ & PA              & $T_{\rm B, obs}$  \\
    (MJD)   & (Jy)        & ($\rm \mu as$) & ($\rm \mu as$)      &($^\circ$)       & ($\rm 10^{11}~K$) \\
    \hline
    $56828$ & $5.6\pm1.0$ & $86.6 \pm0.5$  & $20.0 \pm1.0$       & $-33.15\pm0.32$ & $91.8 \pm18.2$    \\ [0pt]
    $56866$ & $7.2\pm0.6$ & $114.3\pm1.7$  & $64.8 \pm3.4$       & $-56.71\pm0.86$ & $11.2 \pm1.5 $    \\ [0pt]
    $56923$ & $8.9\pm0.5$ & $171.2\pm1.5$  & $76.5 \pm2.9$       & $-62.62\pm0.49$ & $9.9  \pm0.9 $    \\ [0pt]
    $56976$ & $2.3\pm0.4$ & $295.6\pm1.8$  & $57.4 \pm3.5$       & $-56.21\pm0.34$ & $4.6  \pm0.9 $    \\ [0pt]
    $56996$ & $4.4\pm0.2$ & $281.4\pm1.2$  & $65.8 \pm2.4$       & $-61.90\pm0.25$ & $6.6  \pm0.6 $    \\ [0pt]
    $57020$ & $4.5\pm0.3$ & $308.2\pm2.5$  & $106.8\pm4.9$       & $-64.43\pm0.46$ & $2.6  \pm0.3 $    \\ [0pt]
    $57067$ & $7.6\pm0.4$ & $379.0\pm2.1$  & $106.7\pm4.3$       & $-68.15\pm0.32$ & $4.4  \pm0.4 $    \\ [0pt]
    $57123$ & $8.9\pm0.7$ & $428.4\pm3.9$  & $125.7\pm7.7$       & $-68.27\pm0.52$ & $3.7  \pm0.5 $    \\ [0pt]
    $57153$ & $5.2\pm0.3$ & $515.4\pm1.4$  & $94.2 \pm2.8$       & $-68.39\pm0.15$ & $3.9  \pm0.3 $    \\ [0pt]
    $57182$ & $5.8\pm0.4$ & $544.1\pm1.2$  & $54.8 \pm2.4$       & $-73.13\pm0.12$ & $12.6 \pm1.4 $    \\ [0pt]
    $57205$ & $5.4\pm0.4$ & $568.4\pm1.0$  & $52.4 \pm1.9$       & $-71.86\pm0.10$ & $12.9 \pm1.3 $    \\ [0pt]
    $57235$ & $6.3\pm0.4$ & $583.8\pm0.8$  & $45.2 \pm1.7$       & $-73.73\pm0.08$ & $20.3 \pm1.9 $    \\ [0pt]
    $57287$ & $6.1\pm0.5$ & $595.0\pm0.4$  & $19.2 \pm0.9$       & $-74.58\pm0.04$ & $108.6\pm12.8$    \\ [0pt]
    $57361$ & $4.9\pm0.3$ & $614.5\pm1.1$  & $53.2 \pm2.2$       & $-75.79\pm0.10$ & $11.4 \pm1.2 $    \\ [0pt]
    $57388$ & $4.4\pm0.3$ & $612.3\pm0.9$  & $47.7 \pm1.8$       & $-76.90\pm0.09$ & $12.5 \pm1.2 $    \\ [0pt]
    $57418$ & $4.5\pm0.3$ & $608.5\pm1.0$  & $46.2 \pm2.0$       & $-75.92\pm0.10$ & $14.0 \pm1.6 $    \\ [0pt]
    $57465$ & $3.6\pm0.3$ & $603.1\pm0.9$  & $37.4 \pm1.8$       & $-77.29\pm0.09$ & $17.0 \pm2.2 $    \\ [0pt]
    $57500$ & $4.2\pm0.5$ & $598.9\pm1.7$  & $53.6 \pm3.5$       & $-79.61\pm0.17$ & $9.7  \pm1.6 $    \\ [0pt]
    $57549$ & $5.0\pm0.5$ & $550.4\pm3.3$  & $95.7 \pm6.6$       & $-83.60\pm0.34$ & $3.6  \pm0.6 $    \\ [0pt]
    \hline \multicolumn{6}{l}{* Note$\colon$The uncertainties are estimated following \citet{fomalont1999}.} \end{tabular} \end{table}
    
%%%%%%%%%%%%%%%%%%%% Result %%%%%%%%%%%%%%%%%%%%
\section{Results} \label{sec4:Result}

\subsection{Double SSA Spectra} \label{sec4.1:Double SSA Spectra}
As described in Section \ref{sec3.2:SSA Spectra}, we identified two individual SSA spectra from the source spectrum. As summarized in the Table \ref{table:3:SSA Parameters}, $S_{\rm m}$ and $\nu_{\rm m}$ of the LSS continuously decreased until the  $\gamma_{2014}$ flare occurred. Therefore, the LSS is likely to be in an adiabatic expansion phase according to the shock-in-jet model \citep{marscher&gear1985}. After the $\gamma_{2014}$ flare, $S_{\rm m}$ and $\nu_{\rm m}$ suddenly increased by up to an order of magnitude. Meanwhile, the optically thin index flattened. These changes could occur if a plasma population were injected into the jet in that period. \citet{liodakis2020} found that jet component K14 was ejecting from the 43 GHz radio core after $\gamma_{2014}$ flare. 

At the HSS, we found a continuous increase in $S_{\rm m}$ until the $\gamma_{2014}$ flare. In some epochs, the obtained $\alpha_{\rm thin}$ was close to 0 (e.g., $\alpha_{\rm thin}\approx-0.1$), implying a possible convolution of two or multiple SSA components at high frequency.

\subsection{Location of SSA Emitting Regions} \label{sec4.2:SSA Locations}        
As shown in Figure \ref{fig:6:Core-SSA}, the time variation of the HSS-inferred flux density at 43 GHz closely follows that of the 43 GHz core in the time range of January 2013 -- December 2014, indicating that the HSS emitting region is the 43 GHz core. The higher flux density of K14 than that of the core also represents the observed flux densities of the LSS and the HSS at 43 GHz.

On the other hand, the LSS-inferred flux density at 43 GHz follows the flux density of the remaining jet components. At 5 GHz, the flux variations of the LSS marginally follow that of the 5 GHz core. As the core flux density (e.g., $\rm \sim2.6~Jy$ in May 2013) is similar to that ($\rm \sim2.7~Jy$) of the HSS at 43 GHz, the flux density (e.g., less than 2 Jy) of the remaining jet components corresponds to that (e.g., less than 1 Jy) of the LSS at 43 GHz. 

Between January 2013 and June 2014, component C was the brightest jet component at 43 GHz, except for the core. In this period, we found a decrease in flux density of component C (e.g., from 1.5~Jy to 0.5~Jy) and the 5 GHz core (from 5.5 Jy to 3.3 Jy) until the $\gamma_{2014}$. A similar decline is seen in the $S_{\rm m}$ of the LSS (from 2.5 Jy to 1.0 Jy, see Table~\ref{table:3:SSA Parameters}), supporting that component C is the LSS emitting region. We note that we found no significant decrease in the flux density of jet components at 5 GHz. After the $\gamma_{2014}$ flare, the flux density of K14 followed the LSS-inferred flux density at 43 GHz. However, note that there are only three epochs after the flare.

Component C exhibits an optically thin spectrum (Figure~\ref{fig:8:Spectra_C}), and the obtained spectrum is similar to the LSS in 2013 (blue dashed line in Figure \ref{fig:4:SSA_Spectra}). We note that the observing date at 15 GHz differs from other frequencies, but the flux density of component C observed in July 2013 ($\rm \sim1.8~Jy$) is similar to that observed in February 2013 ($\rm \sim2.3~Jy$), keeping the optically thin spectrum. In 2016, after the $\gamma_{2014}$ flare, the flux density of component C increased at all three frequencies. Component C had a $\nu_{\rm m}$ at $\sim$22 GHz in January 2016. Meanwhile, in June 2016, the spectrum became optically thin again. These results indicate that the spectral properties of component C vary with time and indicate that a quasi-stationary component can be an SSA-emitting region.

Based on the core-shift effect, the estimated mean offset between the cores at 43 and 5 GHz is $\rm \sim0.55~mas$, which corresponds to component C \citep[$\rm0.45-0.7~mas$,][]{jorstad2005, jorstad2017, weaver2022}. Therefore, component C is considered a core-like component at 5 GHz, given the core-shift effect and the spectrum.

The above results indicate that the HSS emitting region is at the 43 GHz core. The LSS emitting regions are$\colon$component C before the $\gamma_{2014}$, and K14 before the $\gamma_{2014}$.
    
\subsection{Time-averaged Characteristics in Jet} \label{sec4.3:Time-averaged Characteristics}
Figure~\ref{fig:10:Stacked_Pol_Map} shows the time-averaged maps at 43 GHz. The obtained linear polarization intensity and DP are higher in component C ($\rm \sim200~mJy/beam$, $\sim8\%$) than at the core ($\rm \sim30~mJy/beam$, $\sim1\%$). Those polarimetric characteristics are seen in most individual epochs, indicating that de-polarization of the core by variability is unlikely to bias the stacked result. Moreover, when the core region showed high linear polarization intensity and DP, the moving knot K14 was ejecting from the core, implying that the ejecting component caused these polarimetric characteristics in the core. Typically, in every individual map, the DPs of core and component C were $\sim$~1~$\%$ and $\sim$~10~$\%$, respectively.

Alternatively, the de-polarization in the core can occur due to its complex structure. However, we estimated magnetic field strengths for both the LSS and the HSS (Figure \ref{fig:5:B_Field}) and found that the LSS emitting region before the $\gamma_{2014}$ (i.e.,  component C) is magnetically dominated ($B_{\rm SSA, LSS} \gtrsim B_{\rm EQ, LSS}$). Meanwhile, the HSS emitting region (i.e., the core) is dominated by kinetic energy density rather than magnetic energy density ($B_{\rm SSA, HSS} \lesssim B_{\rm EQ, HSS}$). Therefore, it is possible that the difference in polarization intensity and DP between the core and component C can be attributed to the difference between their intrinsic magnetic field strength.

In the stacked map, component C shows EVPA (black bars) parallel with the jet direction. These parallel EVPA distributions are seen in most of the VLBA observations, as far as component C is detected reliably (i.e., except for the epochs when component C was fading out). Meanwhile, EVPA distribution in the core region has an angle perpendicular to the jet direction, although the EVPA of the core region varies in time. However, note that the EVPA distribution can vary by viewing angle due to a projection effect of magnetic fields \cite[see][]{cawthorne2006, cawthorne2013, fuentes2018}.

\section{Discussion} \label{sec5:Discussion}

\subsection{Interpretations of Magnetic Field Strength and Brightness Temperature} \label{sec5.1:B Field Interpretation}
The estimated B-field strength ($B_{\rm SSA, LSS}$) of component C, before the $\gamma_{2014}$ flare, implies that the region is under near equipartition ($B_{\rm SSA, LSS} \approx B_{\rm EQ, LSS}$) condition in May 2013 (the first epoch), and is dominated by the magnetic field energy density ($B_{\rm SSA, LSS} > B_{\rm EQ, LSS}$) in the other epochs. In an alternative way, the magnetic dominance in component C can be investigated using the $T_{\rm B}$ estimates from the physical parameters of component C (i.e., the LSS, see Section~\ref{sec3.7:Brightness Temperature}). In Table~\ref{table:3:SSA Parameters}, we summarized the obtained parameters, including the  $T_{\rm B, obs}$ of the two SSA emitting regions. Before the $\gamma_{2014}$ flare, the observed brightness temperatures $T_{\rm B, obs}$ were estimated to be $(4-16)\times10^{10}~{\rm K}$. Adopting the Doppler factor $\delta_{\rm LSS}=20.3\pm1.8$ (see Section~\ref{sec3.3:B Field Estimation}), the intrinsic brightness temperatures of component C were estimated to be $T_{\rm B, int} = (1+z) \, T_{\rm B, obs} / \delta_{\rm LSS} \approx (4-15)\times10^{9}~{\rm K}$, which are below the equipartition limit, implying a possible magnetic dominance in component C.

Component C could  exhibit magnetic dominance if the electrons effectively dissipate energy, resulting in the production of $\gamma$-rays. The inverse-Compton scattering may be one of the significant cooling mechanisms. Indeed, we found that the observed brightness temperature of K14 increases during MJD~57123 -- 57287, from $\sim4\times 10^{11}~{\rm K}$ to $\sim1\times 10^{13}~{\rm K}$, peaking on MJD~57287 when K14 reaches component C ($\rm \sim0.6~mas$, a month after the $\gamma_{2015}$ flare). By adopting the typical Doppler factor of 3C~454.3 as $\delta = 28.8\pm2.2$ \citep{weaver2022},  the intrinsic brightness temperature around the $\gamma_{2015}$ flare was computed to be $T_{\rm B, int} \approx (7.0\pm1.0) \times 10^{11}~{\rm K}$, which is close to the inverse-Compton limit. This indicates the inverse-Compton catastrophe, i.e., the efficient energy loss of the electrons (and hence the decrease in particle energy density) to $\gamma$-ray production in component C. The angular size reduction of component K14 around $\gamma_{2015}$ suggests  compression within the jet, possibly due to a recollimation shock.
Alternatively, particle energy loss efficiency to $\gamma$-ray production in component C can be computed using spectral energy distribution (SED). By assuming the high-energy peak of the SED is related to the inverse-Compton scattering, and by using the ratio of the two peaks of the humps in SED and magnetic field strength, the timescale of energy losses of those electrons can be calculated \citep{weaver2020, marscher2022}. Further analysis will be presented in future works.

The lower measurements of $B_{\rm EQ, LSS}$ than $B_{\rm EQ, HSS}$ can be explained if we assume that the LSS emitting region is downstream of the 43 GHz core. In that case, the radiation cooling makes particles lose energy as it moves downstream. In Section~\ref{sec4.2:SSA Locations}, we determined that the SSA emitting regions were the core (HSS, upstream jet) and the component C (LSS, downstream jet) before the flare.

After the $\gamma_{2014}$ flare, however, the estimated $B_{\rm EQ, LSS}$ increased due to the change in both turnover frequency $\nu_{\rm m}$ and interpolated angular size $d_{\rm m}$. Decreased $d_{\rm m}$ after the flare implies that the emission region is either located upstream of the jet than before or experienced collimation. As described in Section \ref{sec3.4:Features of core and C}, the superluminal jet component, K14, emerged from the 43 GHz radio core after the flare. Moreover, component C was fading out until the flare (see Section~\ref{sec4.2:SSA Locations}). These indicate the decrease in $d_{\rm m}$ is attributed to component K14, whose location is a more upstream region in the jet than component C. We obtained similar $B_{\rm EQ}$ between the LSS and the HSS. Given that the HSS emitting region is the 43 GHz core, the similar measurements of $B_{\rm EQ}$ after the flare also support attributing component K14 to the LSS.

\subsection{Possible Recollimation Shock at the Quasi-Stationary Component} \label{sec5.2:Recollimation Shock}
\citet{gomez1999} suggested that component C originated from a recollimation shock in the jet. To further investigate whether a recollimation shock is in component C, we performed a jet ridgeline analysis using the VLBA 43 GHz and 86 GHz maps observed in September 2015 (see Section~\ref{sec3.5:Ridgeline}). The jet ridgeline analysis shows a local minimum of the jet width in component C ($\sim0.6~{\rm mas}$, or $\sim 6\times10^{5}~r_{\rm g}$), in which the jet width is smaller than the estimated jet geometry (see Appendix~\ref{sec:A:Jet_Geo}, and Figure~\ref{fig:A1:Jet_Geo}). Beyond component C, the jet width expands again. A smaller jet width than the estimated jet geometry implies there is an overpressured region in component C. In the previous studies on recollimation shocks, a similar decrease in jet width was identified$\colon$1H~0323+342 \citep{doi2018, hada2018}, 3C~111 \citep{beuchert2018}, and BL Lac \citep{casadio2021}.

\begin{figure} \includegraphics[width=\columnwidth]{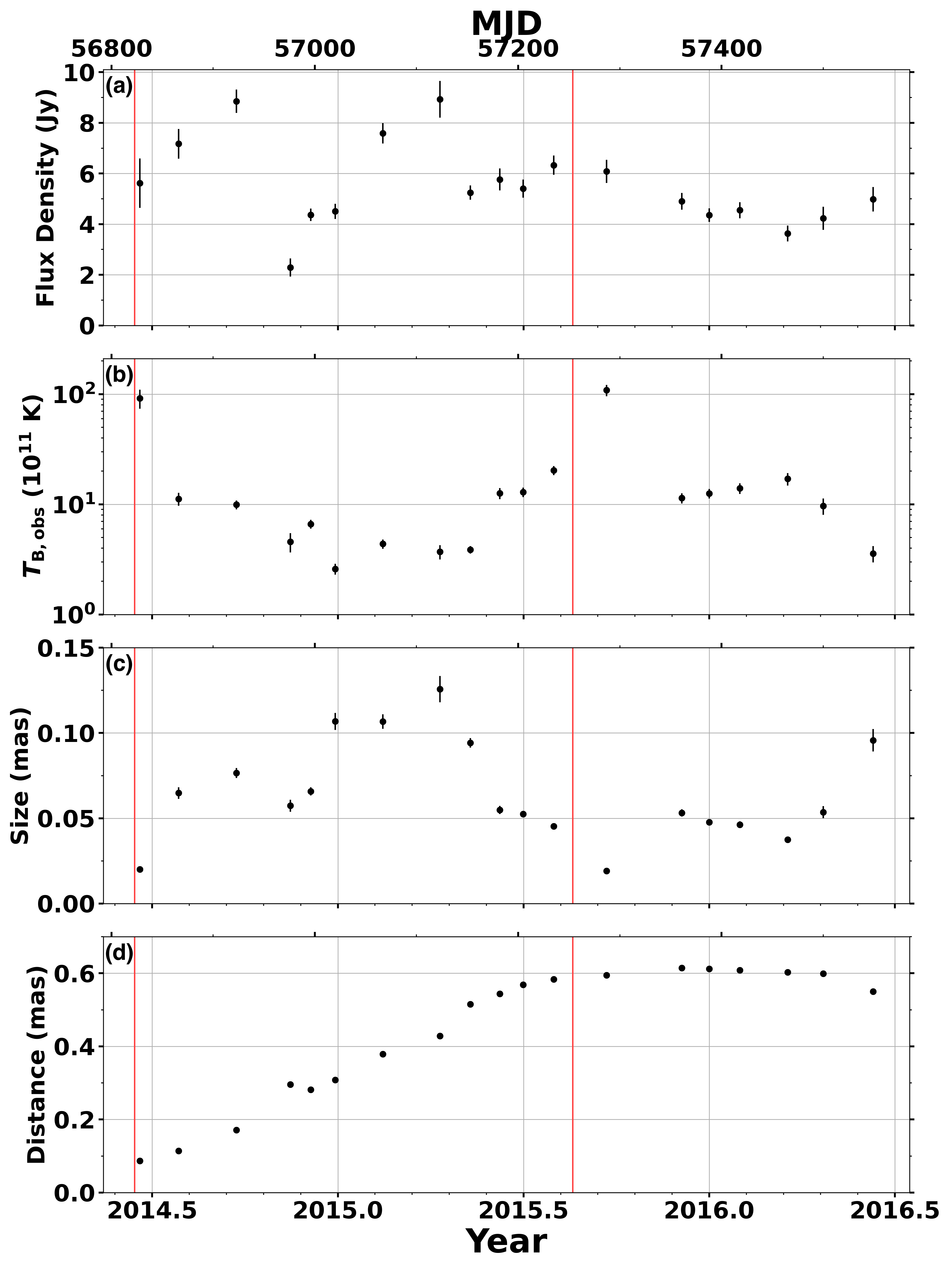}
    \caption{Time variation of flux density (panel a), brightness temperature (panel b), angular size (panel c), and radial distance from the 43 GHz radio core (panel d) of the jet component K14. The two red solid vertical lines indicate $\gamma$-ray flares in June 2014 and August 2015. Note that there are two large decreases in flux density, as the K14 is divided into two components.}
    \label{fig:11:K14} \end{figure}

Figure~\ref{fig:11:K14} shows the variations (flux density, observed brightness temperature, angular size, and distance from the core) of component K14 about two years from its ejection. In the Figure, as component K14 approaches component C $(\sim 0.6~{\rm mas})$, the flux density increases while the angular size decreases, which is clearly seen after the second separation of component K14 (between $\rm MJD~57153-57287$). The decrease in the angular size may imply compression in the jet as component K14 approaches component C. In the same period, we found that the linear polarization intensity at component C also increased from $\rm \sim450~mJy/beam$ to $\rm \sim750~mJy/beam$ around the $\gamma_{2015}$ flare $(\rm MJD~57235-57287)$. The variability in the total flux density of component K14 in this work is similar to that in \citep[][see Figure 7 in their work]{liodakis2020}. Similarly, \citet{jorstad2010} found another re-brightening of jet component passing component C in 3C~454.3. The re-brightening behavior of a component passing a recollimation shock has been reported in previous studies \citep{fromm2013a, beuchert2018, hada2018}.
Based on the observed jet width decrease and re-brightening behavior, we suggest the presence of a recollimation shock in the quasi-stationary component C. A more sensitive VLBI observation at high frequencies (e.g., 86 or 230 GHz) would be
required to clarify the nature of component C in future work.

%%%%%%%%%%%%%%%%%%%% Summary %%%%%%%%%%%%%%%%%%%%
\section{Summary} \label{sec6:Summary}
This paper presents results on 3C~454.3 using radio multi-wavelength data ($\rm 3-340~GHz$). Using the data, we found two peaks in the radio spectra. The first peak is observed at relatively low frequencies of $\rm 3-37~GHz$ (LSS), and the second peak appears at high frequencies of $\rm 55-124~GHz$ (HSS).
We compared variations in the flux density of SSA-inferred and VLBI measurements to investigate emitting regions of the LSS and the HSS. We found that the flux variation of the HSS closely follows that of the 43 GHz radio core. The flux variability of the LSS marginally matches with the remaining jet components. Moreover, at 5 GHz, the LSS-inferred flux variation shows a similar trend with the 5 GHz core. These results imply that the HSS emitting region is located in the 43 GHz radio core, and the emission region for the LSS might be near the 5 GHz core. From the previous works on core-shift effect, a mean distance offset of $\rm \sim0.55~mas$ was found between the cores at 43 GHz and 5 GHz. The distance corresponds to that of the quasi-stationary component C from the core at 43 GHz \citep[$\rm 0.45 \sim 0.70~mas$, ][]{jorstad2005, jorstad2017, weaver2022}.

We investigated the spectrum of component C using VLBA observations at three frequencies. From the spectral analysis, we found that component C showed an optically thin spectrum before the $\gamma_{2014}$ flare and the SSA spectrum with $\rm \nu_{m}\approx22~GHz$ in 2016. The obtained spectrum in 2013 was consistent with the LSS model. Furthermore, we observed a gradual decrease in the flux density of component C until the occurrence of the $\gamma_{2014}$ flare, as seen in the LSS model (Table~\ref{table:3:SSA Parameters}, and Figure~\ref{fig:4:SSA_Spectra}). In addition, we discovered the emergence of a new jet component, K14, which exhibited a higher flux density than the core. This component was ejected from the 43 GHz core subsequent to the flare. Based on these observations, we can conclude that the LSS feature can be attributed to component C prior to the flare and to component K14 following the flare.

In August 2015, during the $\gamma_{2015}$ flare, the moving component K14 reached a distance of approximately 0.6 mas from the core. As K14 approached component C, there was a noticeable re-brightening observed in both the total flux density (increasing from approximately 5 Jy to 6 Jy) and the linearly polarized intensity (rising from around 450 mJy/beam to 750 mJy/beam). Furthermore, component K14 underwent a diminution in its angular size during this period.

By adopting a typical Doppler factor, $\delta=28.8\pm2.2$ \citep{weaver2022} for the source, we found that the intrinsic brightness temperature $T_{\rm B, int}\approx(7.0\pm1.0)\times10^{11}~{\rm K}$ of K14 is comparable with the inverse-Compton limit \citep[$T_{\rm IC}\approx10^{12}~{\rm K}$,][]{kellermann1969_radiosources} in $\rm MJD~57287$. This finding suggests that the interaction between components K14 and C is linked to the $\gamma_{2015}$ flare. Notably, the $\gamma$-ray production in component C potentially leads to a decrease in particle energy density, resulting in component C becoming a magnetically dominated region.

The higher magnetic field strength ($B_{\rm SSA}$) than that of the equipartition condition ($B_{\rm EQ}$) supports the magnetic dominance in component C (Table \ref{table:4:B_Strength}). We also investigated $T_{\rm B, int}$ of the LSS using its parameters in Table \ref{table:3:SSA Parameters} and $\delta_{\rm LSS} = 20.3 \pm 1.8$ (see Section \ref{sec3.3:B Field Estimation}). The calculated values were lower than the equipartition condition \citep[$T_{\rm EQ} \approx 5 \times 10^{10}~{\rm K}$,][]{readhead1994} before the $\gamma_{2014}$ flare, confirming the magnetic dominance in component C.

The features related to component C, re-brightening behavior, magnetic dominance, and the angular size reduction in K14 can be explained if component C is an overpressured region, possibly due to a recollimation shock. The jet ridgeline analysis showed hints of recollimation near component C (see Section~\ref{sec3.5:Ridgeline}). Moreover, the estimated jet width in component C is smaller than the jet geometry~(see Section~\ref{sec:A:Jet_Geo}), implying an overpressured region.

Time-averaged linear polarization characteristics were investigated by stacking the 38 VLBA 43 GHz observations observed in the period of January 2013 -- December 2016. In the stacked map, component C shows high polarization intensity ($\sim200~{\rm mJy/beam}$) and DP ($\sim8~\%$).
All of the characteristics related to component C may suggest a presence of recollimation shock in the jet of 3C~454.3.

%%%%%%%%%%%%%%%%%%%% Acknowledgements %%%%%%%%%%%%%%%%%%%%
\section*{Acknowledgements}
We thank the anonymous reviewer for valuable comments and suggestions that helped to improve the paper. We are grateful to the staff of the KVN who helped to operate the array and to correlate the data. The KVN is a facility operated by the KASI (Korea Astronomy and Space Science Institute). The KVN observations and correlations are supported through high-speed network connections among the KVN sites provided by the KREONET (Korea Research Environment Open NETwork), which is managed and operated by the KISTI (Korea Institute of Science and Technology Information). We genuinely thank Prof. Alan P. Marscher and Dr. Svetlana G. Jorstad for their valuable comments on this work. This research made use of data from the OVRO 40-m monitoring program \citep{richards2011},  supported by private funding from the California Institute of Technology and the Max Planck Institute for Radio Astronomy, and by NASA grants NNX08AW31G, NNX11A043G, and NNX14AQ89G and NSF grants AST-0808050 and AST- 1109911. This paper has used data from the MARMOT program that was a key-science program running at CARMA. The data from this program are provided by Dr. Talvikki Hovatta. The Submillimeter Array is a joint project between the Smithsonian Astrophysical Observatory and the Academia Sinica Institute of Astronomy and Astrophysics and is funded by the Smithsonian Institution and the Academia Sinica. We recognize that Maunakea is a culturally important site for the indigenous Hawaiian people; we are privileged to study the cosmos from its summit. This paper makes use of the following ALMA data: ADS/JAO.ALMA$\#$2011.0.01234.S. ALMA is a partnership of ESO (representing its member states), NSF (USA) and NINS (Japan), together with NRC (Canada), MOST and ASIAA (Taiwan), and KASI (Republic of Korea), in cooperation with the Republic of Chile. The Joint ALMA Observatory is operated by ESO, AUI/NRAO and NAOJ.  This research has made use of data from the MOJAVE database that is maintained by the MOJAVE team \citep{lister2018}. This study makes use of VLBA data from the VLBA-BU Blazar Monitoring Program (BEAM-ME and VLBA-BU-BLAZAR; http://www.bu.edu/blazars/BEAM-ME.html), funded by NASA through the Fermi Guest Investigator Program. The VLBA is an instrument of the National Radio Astronomy Observatory. The National Radio Astronomy Observatory is a facility of the National Science Foundation operated by Associated Universities, Inc. This research made use of Astropy,\footnote{\url{http://www.astropy.org}} a community-developed core Python package for Astronomy \citep{astropy:2013, astropy:2018}. This work was supported by the National Research Foundation of Korea (NRF) grant funded by the Korea government (MIST) (2020R1A2C2009003). JYK acknowledges support from the National Research Foundation of Korea (grant no. 2022R1C1C1005255).

%%%%%%%%%%%%%%%%%%%% DATA AVAILABILITY %%%%%%%%%%%%%%%%%%%%
\section*{DATA AVAILABILITY}
 The data underlying this article will be shared on reasonable request to the corresponding author. 
 The CARMA monitoring data at 95 GHz can be obtained from the MARMOT website at https://sites.astro.caltech.edu/marmot/index.php. Please contact Talvikki Hovatta (talvikki.hovatta at aalto.fi) regarding the CARMA data. The 15 GHz OVRO data may be available on request to the OVRO 40 m collaboration. The SMA data are available at http://sma1.sma.hawaii.edu/callist/callist.html. Questions regarding the data and data policies should be addressed to Mark Gurwell (mgurwell at cfa.harvard.edu). The ALMA data are available at the ALMA Calibrator Source Catalogue, found at https://almascience.eso.org/sc/. The 15 GHz VLBA data from the MOJAVE monitoring program can be obtained at https://www.cv.nrao.edu/MOJAVE/. The 43 GHz VLBA data from the VLBA-BU-BLAZAR program are available at https://www.bu.edu/blazars/VLBAproject.html. The GMVA 86 GHz data can be obtained by contacting Alan. P. Marscher (marscher at bu.edu).

%%%%%%%%%%%%%%%%%%%% REFERENCES %%%%%%%%%%%%%%%%%%%%
\bibliographystyle{mnras} \bibliography{CR_paper_2col}

\begin{thebibliography}{}
\makeatletter
\relax
\def\mn@urlcharsother{\let\do\@makeother \do\$\do\&\do\#\do\^\do\_\do\%\do\~}
\def\mn@doi{\begingroup\mn@urlcharsother \@ifnextchar [ {\mn@doi@}
  {\mn@doi@[]}}
\def\mn@doi@[#1]#2{\def\@tempa{#1}\ifx\@tempa\@empty \href
  {http://dx.doi.org/#2} {doi:#2}\else \href {http://dx.doi.org/#2} {#1}\fi
  \endgroup}
\def\mn@eprint#1#2{\mn@eprint@#1:#2::\@nil}
\def\mn@eprint@arXiv#1{\href {http://arxiv.org/abs/#1} {{\tt arXiv:#1}}}
\def\mn@eprint@dblp#1{\href {http://dblp.uni-trier.de/rec/bibtex/#1.xml}
  {dblp:#1}}
\def\mn@eprint@#1:#2:#3:#4\@nil{\def\@tempa {#1}\def\@tempb {#2}\def\@tempc
  {#3}\ifx \@tempc \@empty \let \@tempc \@tempb \let \@tempb \@tempa \fi \ifx
  \@tempb \@empty \def\@tempb {arXiv}\fi \@ifundefined
  {mn@eprint@\@tempb}{\@tempb:\@tempc}{\expandafter \expandafter \csname
  mn@eprint@\@tempb\endcsname \expandafter{\@tempc}}}

\bibitem[\protect\citeauthoryear{{Algaba} et~al.,}{{Algaba}
  et~al.}{2018}]{algaba2018}
{Algaba} J.-C.,  et~al., 2018, \mn@doi [\apj] {10.3847/1538-4357/aac2e7}, \href
  {https://ui.adsabs.harvard.edu/abs/2018ApJ...859..128A} {859, 128}

\bibitem[\protect\citeauthoryear{{Amaya-Almaz{\'a}n}, {Chavushyan}  \&
  {Pati{\~n}o-{\'A}lvarez}}{{Amaya-Almaz{\'a}n}
  et~al.}{2021}]{amaya_almazan2021}
{Amaya-Almaz{\'a}n} R.~A.,  {Chavushyan} V.,   {Pati{\~n}o-{\'A}lvarez} V.~M.,
  2021, \mn@doi [\apj] {10.3847/1538-4357/abc689}, \href
  {https://ui.adsabs.harvard.edu/abs/2021ApJ...906....5A} {906, 5}

\bibitem[\protect\citeauthoryear{{Angelakis} et~al.,}{{Angelakis}
  et~al.}{2019}]{angelakis2019}
{Angelakis} E.,  et~al., 2019, \mn@doi [\aap] {10.1051/0004-6361/201834363},
  \href {https://ui.adsabs.harvard.edu/abs/2019A&A...626A..60A} {626, A60}

\bibitem[\protect\citeauthoryear{{Astropy Collaboration} et~al.,}{{Astropy
  Collaboration} et~al.}{2013}]{astropy:2013}
{Astropy Collaboration} et~al., 2013, \mn@doi [\aap]
  {10.1051/0004-6361/201322068}, \href
  {https://ui.adsabs.harvard.edu/abs/2013A&A...558A..33A} {558, A33}

\bibitem[\protect\citeauthoryear{{Astropy Collaboration} et~al.,}{{Astropy
  Collaboration} et~al.}{2018}]{astropy:2018}
{Astropy Collaboration} et~al., 2018, \mn@doi [\aj] {10.3847/1538-3881/aabc4f},
  \href {https://ui.adsabs.harvard.edu/abs/2018AJ....156..123A} {156, 123}

\bibitem[\protect\citeauthoryear{{Beuchert} et~al.,}{{Beuchert}
  et~al.}{2018}]{beuchert2018}
{Beuchert} T.,  et~al., 2018, \mn@doi [\aap] {10.1051/0004-6361/201731952},
  \href {https://ui.adsabs.harvard.edu/abs/2018A&A...610A..32B} {610, A32}

\bibitem[\protect\citeauthoryear{{B{\"o}ttcher} et~al.,}{{B{\"o}ttcher}
  et~al.}{2007}]{bottcher2007}
{B{\"o}ttcher} M.,  et~al., 2007, \mn@doi [\apj] {10.1086/522583}, \href
  {https://ui.adsabs.harvard.edu/abs/2007ApJ...670..968B} {670, 968}

\bibitem[\protect\citeauthoryear{{B{\"o}ttcher}, {Reimer}, {Sweeney}  \&
  {Prakash}}{{B{\"o}ttcher} et~al.}{2013}]{bottcher2013}
{B{\"o}ttcher} M.,  {Reimer} A.,  {Sweeney} K.,   {Prakash} A.,  2013, \mn@doi
  [\apj] {10.1088/0004-637X/768/1/54}, \href
  {https://ui.adsabs.harvard.edu/abs/2013ApJ...768...54B} {768, 54}

\bibitem[\protect\citeauthoryear{{Buson}}{{Buson}}{2014}]{gamma2014}
{Buson} S.,  2014, The Astronomer's Telegram, \href
  {https://ui.adsabs.harvard.edu/abs/2014ATel.6236....1B} {6236, 1}

\bibitem[\protect\citeauthoryear{{Casadio} et~al.,}{{Casadio}
  et~al.}{2019}]{casadio2019}
{Casadio} C.,  et~al., 2019, \mn@doi [\aap] {10.1051/0004-6361/201834519},
  \href {https://ui.adsabs.harvard.edu/abs/2019A&A...622A.158C} {622, A158}

\bibitem[\protect\citeauthoryear{{Casadio} et~al.,}{{Casadio}
  et~al.}{2021}]{casadio2021}
{Casadio} C.,  et~al., 2021, \mn@doi [\aap] {10.1051/0004-6361/202039616},
  \href {https://ui.adsabs.harvard.edu/abs/2021A&A...649A.153C} {649, A153}

\bibitem[\protect\citeauthoryear{{Cawthorne}}{{Cawthorne}}{2006}]{cawthorne2006}
{Cawthorne} T.~V.,  2006, \mn@doi [\mnras] {10.1111/j.1365-2966.2006.10019.x},
  \href {https://ui.adsabs.harvard.edu/abs/2006MNRAS.367..851C} {367, 851}

\bibitem[\protect\citeauthoryear{{Cawthorne}, {Jorstad}  \&
  {Marscher}}{{Cawthorne} et~al.}{2013}]{cawthorne2013}
{Cawthorne} T.~V.,  {Jorstad} S.~G.,   {Marscher} A.~P.,  2013, \mn@doi [\apj]
  {10.1088/0004-637X/772/1/14}, \href
  {https://ui.adsabs.harvard.edu/abs/2013ApJ...772...14C} {772, 14}

\bibitem[\protect\citeauthoryear{{Chamani}, {Savolainen}, {Ros}, {Kovalev},
  {Wiik}, {L{\"a}hteenm{\"a}ki}, {Tornikoski}  \& {Tammi}}{{Chamani}
  et~al.}{2022}]{chamani2022}
{Chamani} W.,  {Savolainen} T.,  {Ros} E.,  {Kovalev} Y.~Y.,  {Wiik} K.,
  {L{\"a}hteenm{\"a}ki} A.,  {Tornikoski} M.,   {Tammi} J.,  2022, arXiv
  e-prints, \href {https://ui.adsabs.harvard.edu/abs/2022arXiv220913301C} {p.
  arXiv:2209.13301}

\bibitem[\protect\citeauthoryear{{Dermer}, {Finke}, {Krug}  \&
  {B{\"o}ttcher}}{{Dermer} et~al.}{2009}]{dermer2009}
{Dermer} C.~D.,  {Finke} J.~D.,  {Krug} H.,   {B{\"o}ttcher} M.,  2009, \mn@doi
  [\apj] {10.1088/0004-637X/692/1/32}, \href
  {https://ui.adsabs.harvard.edu/abs/2009ApJ...692...32D} {692, 32}

\bibitem[\protect\citeauthoryear{{Doi}, {Hada}, {Kino}, {Wajima}  \&
  {Nakahara}}{{Doi} et~al.}{2018}]{doi2018}
{Doi} A.,  {Hada} K.,  {Kino} M.,  {Wajima} K.,   {Nakahara} S.,  2018, \mn@doi
  [\apjl] {10.3847/2041-8213/aabae2}, \href
  {https://ui.adsabs.harvard.edu/abs/2018ApJ...857L...6D} {857, L6}

\bibitem[\protect\citeauthoryear{{Fomalont}}{{Fomalont}}{1999}]{fomalont1999}
{Fomalont} E.~B.,  1999, in {Taylor} G.~B.,  {Carilli} C.~L.,   {Perley} R.~A.,
   eds,  Astronomical Society of the Pacific Conference Series Vol. 180,
  Synthesis Imaging in Radio Astronomy II. p.~301

\bibitem[\protect\citeauthoryear{{Foreman-Mackey}, {Hogg}, {Lang}  \&
  {Goodman}}{{Foreman-Mackey} et~al.}{2013}]{emcee2013}
{Foreman-Mackey} D.,  {Hogg} D.~W.,  {Lang} D.,   {Goodman} J.,  2013, \mn@doi
  [\pasp] {10.1086/670067}, \href
  {https://ui.adsabs.harvard.edu/abs/2013PASP..125..306F} {125, 306}

\bibitem[\protect\citeauthoryear{{Fromm} et~al.,}{{Fromm}
  et~al.}{2013a}]{fromm2013a}
{Fromm} C.~M.,  et~al., 2013a, \mn@doi [\aap] {10.1051/0004-6361/201219913},
  \href {https://ui.adsabs.harvard.edu/abs/2013A&A...551A..32F} {551, A32}

\bibitem[\protect\citeauthoryear{{Fromm}, {Ros}, {Perucho}, {Savolainen},
  {Mimica}, {Kadler}, {Lobanov}  \& {Zensus}}{{Fromm}
  et~al.}{2013b}]{fromm2013b}
{Fromm} C.~M.,  {Ros} E.,  {Perucho} M.,  {Savolainen} T.,  {Mimica} P.,
  {Kadler} M.,  {Lobanov} A.~P.,   {Zensus} J.~A.,  2013b, \mn@doi [\aap]
  {10.1051/0004-6361/201321784}, \href
  {https://ui.adsabs.harvard.edu/abs/2013A&A...557A.105F} {557, A105}

\bibitem[\protect\citeauthoryear{{Fuentes}, {G{\'o}mez}, {Mart{\'\i}}  \&
  {Perucho}}{{Fuentes} et~al.}{2018}]{fuentes2018}
{Fuentes} A.,  {G{\'o}mez} J.~L.,  {Mart{\'\i}} J.~M.,   {Perucho} M.,  2018,
  \mn@doi [\apj] {10.3847/1538-4357/aac091}, \href
  {https://ui.adsabs.harvard.edu/abs/2018ApJ...860..121F} {860, 121}

\bibitem[\protect\citeauthoryear{{Fuhrmann} et~al.,}{{Fuhrmann}
  et~al.}{2016}]{fuhrmann2016}
{Fuhrmann} L.,  et~al., 2016, \mn@doi [\aap] {10.1051/0004-6361/201528034},
  \href {https://ui.adsabs.harvard.edu/abs/2016A&A...596A..45F} {596, A45}

\bibitem[\protect\citeauthoryear{{G{\'o}mez}, {Marscher}  \&
  {Alberdi}}{{G{\'o}mez} et~al.}{1999}]{gomez1999}
{G{\'o}mez} J.-L.,  {Marscher} A.~P.,   {Alberdi} A.,  1999, \mn@doi [\apj]
  {10.1086/307645}, \href
  {https://ui.adsabs.harvard.edu/abs/1999ApJ...522...74G} {522, 74}

\bibitem[\protect\citeauthoryear{{Gurwell}, {Peck}, {Hostler}, {Darrah}  \&
  {Katz}}{{Gurwell} et~al.}{2007}]{gurwell2007}
{Gurwell} M.~A.,  {Peck} A.~B.,  {Hostler} S.~R.,  {Darrah} M.~R.,   {Katz}
  C.~A.,  2007, in {Baker} A.~J.,  {Glenn} J.,  {Harris} A.~I.,  {Mangum}
  J.~G.,   {Yun} M.~S.,  eds,  Astronomical Society of the Pacific Conference
  Series Vol. 375, From Z-Machines to ALMA: (Sub)Millimeter Spectroscopy of
  Galaxies. p.~234

\bibitem[\protect\citeauthoryear{{Hada} et~al.,}{{Hada}
  et~al.}{2018}]{hada2018}
{Hada} K.,  et~al., 2018, \mn@doi [\apj] {10.3847/1538-4357/aac49f}, \href
  {https://ui.adsabs.harvard.edu/abs/2018ApJ...860..141H} {860, 141}

\bibitem[\protect\citeauthoryear{{Homan} et~al.,}{{Homan}
  et~al.}{2021}]{homan2021}
{Homan} D.~C.,  et~al., 2021, \mn@doi [\apj] {10.3847/1538-4357/ac27af}, \href
  {https://ui.adsabs.harvard.edu/abs/2021ApJ...923...67H} {923, 67}

\bibitem[\protect\citeauthoryear{{Jackson} \& {Browne}}{{Jackson} \&
  {Browne}}{1991}]{jackson1991}
{Jackson} N.,  {Browne} I.~W.~A.,  1991, \mn@doi [\mnras]
  {10.1093/mnras/250.2.414}, \href
  {https://ui.adsabs.harvard.edu/abs/1991MNRAS.250..414J} {250, 414}

\bibitem[\protect\citeauthoryear{{Jorstad} et~al.,}{{Jorstad}
  et~al.}{2005}]{jorstad2005}
{Jorstad} S.~G.,  et~al., 2005, \mn@doi [\aj] {10.1086/444593}, \href
  {https://ui.adsabs.harvard.edu/abs/2005AJ....130.1418J} {130, 1418}

\bibitem[\protect\citeauthoryear{{Jorstad} et~al.,}{{Jorstad}
  et~al.}{2010}]{jorstad2010}
{Jorstad} S.~G.,  et~al., 2010, \mn@doi [\apj] {10.1088/0004-637X/715/1/362},
  \href {https://ui.adsabs.harvard.edu/abs/2010ApJ...715..362J} {715, 362}

\bibitem[\protect\citeauthoryear{{Jorstad} et~al.,}{{Jorstad}
  et~al.}{2013}]{jorstad2013}
{Jorstad} S.~G.,  et~al., 2013, \mn@doi [\apj] {10.1088/0004-637X/773/2/147},
  \href {https://ui.adsabs.harvard.edu/abs/2013ApJ...773..147J} {773, 147}

\bibitem[\protect\citeauthoryear{{Jorstad}, {Larionov}, {Mokrushina},
  {Troitsky}  \& {Morozova}}{{Jorstad} et~al.}{2015}]{gamma2015}
{Jorstad} S.,  {Larionov} V.,  {Mokrushina} A.,  {Troitsky} I.,   {Morozova}
  D.,  2015, The Astronomer's Telegram, \href
  {https://ui.adsabs.harvard.edu/abs/2015ATel.7942....1J} {7942, 1}

\bibitem[\protect\citeauthoryear{{Jorstad} et~al.,}{{Jorstad}
  et~al.}{2017}]{jorstad2017}
{Jorstad} S.~G.,  et~al., 2017, \mn@doi [\apj] {10.3847/1538-4357/aa8407},
  \href {https://ui.adsabs.harvard.edu/abs/2017ApJ...846...98J} {846, 98}

\bibitem[\protect\citeauthoryear{{Kang} et~al.,}{{Kang}
  et~al.}{2021}]{kangsc2021}
{Kang} S.,  et~al., 2021, \mn@doi [\aap] {10.1051/0004-6361/202040198}, \href
  {https://ui.adsabs.harvard.edu/abs/2021A&A...651A..74K} {651, A74}

\bibitem[\protect\citeauthoryear{{Kataoka} \& {Stawarz}}{{Kataoka} \&
  {Stawarz}}{2005}]{kataoka&stawarz2005}
{Kataoka} J.,  {Stawarz} {\L}.,  2005, \mn@doi [\apj] {10.1086/428083}, \href
  {https://ui.adsabs.harvard.edu/abs/2005ApJ...622..797K} {622, 797}

\bibitem[\protect\citeauthoryear{{Kellermann} \& {Pauliny-Toth}}{{Kellermann}
  \& {Pauliny-Toth}}{1969}]{kellermann1969_radiosources}
{Kellermann} K.~I.,  {Pauliny-Toth} I.~I.~K.,  1969, \mn@doi [\apjl]
  {10.1086/180305}, \href
  {https://ui.adsabs.harvard.edu/abs/1969ApJ...155L..71K} {155, L71}

\bibitem[\protect\citeauthoryear{{Kemball}, {Diamond}  \&
  {Pauliny-Toth}}{{Kemball} et~al.}{1996}]{kemball1996}
{Kemball} A.~J.,  {Diamond} P.~J.,   {Pauliny-Toth} I.~I.~K.,  1996, \mn@doi
  [\apjl] {10.1086/310088}, \href
  {https://ui.adsabs.harvard.edu/abs/1996ApJ...464L..55K} {464, L55}

\bibitem[\protect\citeauthoryear{{Kim} et~al.,}{{Kim} et~al.}{2022}]{kimsh2022}
{Kim} S.-H.,  et~al., 2022, \mn@doi [\mnras] {10.1093/mnras/stab3473}, \href
  {https://ui.adsabs.harvard.edu/abs/2022MNRAS.510..815K} {510, 815}

\bibitem[\protect\citeauthoryear{{Komatsu} et~al.,}{{Komatsu}
  et~al.}{2009}]{komatsu2009}
{Komatsu} E.,  et~al., 2009, \mn@doi [\apjs] {10.1088/0067-0049/180/2/330},
  \href {https://ui.adsabs.harvard.edu/abs/2009ApJS..180..330K} {180, 330}

\bibitem[\protect\citeauthoryear{{Kovalev}, {Pushkarev}, {Nokhrina}, {Plavin},
  {Beskin}, {Chernoglazov}, {Lister}  \& {Savolainen}}{{Kovalev}
  et~al.}{2020}]{kovalev2020}
{Kovalev} Y.~Y.,  {Pushkarev} A.~B.,  {Nokhrina} E.~E.,  {Plavin} A.~V.,
  {Beskin} V.~S.,  {Chernoglazov} A.~V.,  {Lister} M.~L.,   {Savolainen} T.,
  2020, \mn@doi [\mnras] {10.1093/mnras/staa1121}, \href
  {https://ui.adsabs.harvard.edu/abs/2020MNRAS.495.3576K} {495, 3576}

\bibitem[\protect\citeauthoryear{{Kutkin} et~al.,}{{Kutkin}
  et~al.}{2014}]{kutkin2014}
{Kutkin} A.~M.,  et~al., 2014, \mn@doi [\mnras] {10.1093/mnras/stt2133}, \href
  {https://ui.adsabs.harvard.edu/abs/2014MNRAS.437.3396K} {437, 3396}

\bibitem[\protect\citeauthoryear{{Lee}, {Lobanov}, {Krichbaum}, {Witzel},
  {Zensus}, {Bremer}, {Greve}  \& {Grewing}}{{Lee} et~al.}{2008}]{lee2008}
{Lee} S.-S.,  {Lobanov} A.~P.,  {Krichbaum} T.~P.,  {Witzel} A.,  {Zensus} A.,
  {Bremer} M.,  {Greve} A.,   {Grewing} M.,  2008, \mn@doi [\aj]
  {10.1088/0004-6256/136/1/159}, \href
  {https://ui.adsabs.harvard.edu/abs/2008AJ....136..159L} {136, 159}

\bibitem[\protect\citeauthoryear{{Lee}, {Han}, {Kang}, {Seen}, {Byun}, {Baek},
  {Kim}  \& {Kim}}{{Lee} et~al.}{2013}]{mogaba_lee2013}
{Lee} S.-S.,  {Han} M.,  {Kang} S.,  {Seen} J.,  {Byun} D.-Y.,  {Baek} J.-H.,
  {Kim} S.-W.,   {Kim} J.-S.,  2013, in European Physical Journal Web of
  Conferences. p. 07007 (\mn@eprint {arXiv} {1311.3375}),
  \mn@doi{10.1051/epjconf/20136107007}

\bibitem[\protect\citeauthoryear{{Lee}, {Lobanov}, {Krichbaum}  \&
  {Zensus}}{{Lee} et~al.}{2016}]{lee2016}
{Lee} S.-S.,  {Lobanov} A.~P.,  {Krichbaum} T.~P.,   {Zensus} J.~A.,  2016,
  \mn@doi [\apj] {10.3847/0004-637X/826/2/135}, \href
  {https://ui.adsabs.harvard.edu/abs/2016ApJ...826..135L} {826, 135}

\bibitem[\protect\citeauthoryear{{Lee}, {Lee}, {Hodgson}, {Kim}, {Algaba},
  {Kang}, {Kang}  \& {Kim}}{{Lee} et~al.}{2017}]{leejw2017}
{Lee} J.~W.,  {Lee} S.-S.,  {Hodgson} J.~A.,  {Kim} D.-W.,  {Algaba} J.-C.,
  {Kang} S.,  {Kang} J.,   {Kim} S.~S.,  2017, \mn@doi [\apj]
  {10.3847/1538-4357/aa72f7}, \href
  {https://ui.adsabs.harvard.edu/abs/2017ApJ...841..119L} {841, 119}

\bibitem[\protect\citeauthoryear{{Lee} et~al.,}{{Lee} et~al.}{2020}]{leejw2020}
{Lee} J.~W.,  et~al., 2020, \mn@doi [\apj] {10.3847/1538-4357/abb4e5}, \href
  {https://ui.adsabs.harvard.edu/abs/2020ApJ...902..104L} {902, 104}

\bibitem[\protect\citeauthoryear{{Liodakis} et~al.,}{{Liodakis}
  et~al.}{2020}]{liodakis2020}
{Liodakis} I.,  et~al., 2020, \mn@doi [\apj] {10.3847/1538-4357/abb1b8}, \href
  {https://ui.adsabs.harvard.edu/abs/2020ApJ...902...61L} {902, 61}

\bibitem[\protect\citeauthoryear{{Lister} et~al.,}{{Lister}
  et~al.}{2009}]{lister2009a}
{Lister} M.~L.,  et~al., 2009, \mn@doi [\aj] {10.1088/0004-6256/138/6/1874},
  \href {https://ui.adsabs.harvard.edu/abs/2009AJ....138.1874L} {138, 1874}

\bibitem[\protect\citeauthoryear{{Lister} et~al.,}{{Lister}
  et~al.}{2013}]{lister2013}
{Lister} M.~L.,  et~al., 2013, \mn@doi [\aj] {10.1088/0004-6256/146/5/120},
  \href {https://ui.adsabs.harvard.edu/abs/2013AJ....146..120L} {146, 120}

\bibitem[\protect\citeauthoryear{{Lister} et~al.,}{{Lister}
  et~al.}{2016}]{lister2016}
{Lister} M.~L.,  et~al., 2016, \mn@doi [\aj] {10.3847/0004-6256/152/1/12},
  \href {https://ui.adsabs.harvard.edu/abs/2016AJ....152...12L} {152, 12}

\bibitem[\protect\citeauthoryear{{Lister}, {Aller}, {Aller}, {Hodge}, {Homan},
  {Kovalev}, {Pushkarev}  \& {Savolainen}}{{Lister} et~al.}{2018}]{lister2018}
{Lister} M.~L.,  {Aller} M.~F.,  {Aller} H.~D.,  {Hodge} M.~A.,  {Homan} D.~C.,
   {Kovalev} Y.~Y.,  {Pushkarev} A.~B.,   {Savolainen} T.,  2018, \mn@doi
  [\apjs] {10.3847/1538-4365/aa9c44}, \href
  {https://ui.adsabs.harvard.edu/abs/2018ApJS..234...12L} {234, 12}

\bibitem[\protect\citeauthoryear{{Lister}, {Homan}, {Kellermann}, {Kovalev},
  {Pushkarev}, {Ros}  \& {Savolainen}}{{Lister} et~al.}{2021}]{lister2021}
{Lister} M.~L.,  {Homan} D.~C.,  {Kellermann} K.~I.,  {Kovalev} Y.~Y.,
  {Pushkarev} A.~B.,  {Ros} E.,   {Savolainen} T.,  2021, \mn@doi [\apj]
  {10.3847/1538-4357/ac230f}, \href
  {https://ui.adsabs.harvard.edu/abs/2021ApJ...923...30L} {923, 30}

\bibitem[\protect\citeauthoryear{{Lobanov}}{{Lobanov}}{1998}]{lobanov1998}
{Lobanov} A.~P.,  1998, \aap, \href
  {https://ui.adsabs.harvard.edu/abs/1998A&A...330...79L} {330, 79}

\bibitem[\protect\citeauthoryear{{Marscher}}{{Marscher}}{1983}]{marscher1983}
{Marscher} A.~P.,  1983, \mn@doi [\apj] {10.1086/160597}, \href
  {https://ui.adsabs.harvard.edu/abs/1983ApJ...264..296M} {264, 296}

\bibitem[\protect\citeauthoryear{{Marscher} \& {Gear}}{{Marscher} \&
  {Gear}}{1985}]{marscher&gear1985}
{Marscher} A.~P.,  {Gear} W.~K.,  1985, \mn@doi [\apj] {10.1086/163592}, \href
  {https://ui.adsabs.harvard.edu/abs/1985ApJ...298..114M} {298, 114}

\bibitem[\protect\citeauthoryear{{Marscher} \& {Jorstad}}{{Marscher} \&
  {Jorstad}}{2022}]{marscher2022}
{Marscher} A.~P.,  {Jorstad} S.~G.,  2022, \mn@doi [Universe]
  {10.3390/universe8120644}, \href
  {https://ui.adsabs.harvard.edu/abs/2022Univ....8..644M} {8, 644}

\bibitem[\protect\citeauthoryear{{Mohan} et~al.,}{{Mohan}
  et~al.}{2015}]{mohan2015}
{Mohan} P.,  et~al., 2015, \mn@doi [\mnras] {10.1093/mnras/stv1412}, \href
  {https://ui.adsabs.harvard.edu/abs/2015MNRAS.452.2004M} {452, 2004}

\bibitem[\protect\citeauthoryear{{O'Sullivan} \& {Gabuzda}}{{O'Sullivan} \&
  {Gabuzda}}{2009}]{osullivan&gabuzda2009}
{O'Sullivan} S.~P.,  {Gabuzda} D.~C.,  2009, \mn@doi [\mnras]
  {10.1111/j.1365-2966.2009.15428.x}, \href
  {https://ui.adsabs.harvard.edu/abs/2009MNRAS.400...26O} {400, 26}

\bibitem[\protect\citeauthoryear{{Park} \& {Algaba}}{{Park} \&
  {Algaba}}{2022}]{park2022_review}
{Park} J.,  {Algaba} J.~C.,  2022, \mn@doi [Galaxies]
  {10.3390/galaxies10050102}, \href
  {https://ui.adsabs.harvard.edu/abs/2022Galax..10..102P} {10, 102}

\bibitem[\protect\citeauthoryear{{Pauliny-Toth}, {Porcas}, {Zensus},
  {Kellermann}, {Wu}, {Nicholson}  \& {Mantovani}}{{Pauliny-Toth}
  et~al.}{1987}]{pauliny-toth1987}
{Pauliny-Toth} I.~I.~K.,  {Porcas} R.~W.,  {Zensus} J.~A.,  {Kellermann} K.~I.,
   {Wu} S.~Y.,  {Nicholson} G.~D.,   {Mantovani} F.,  1987, \mn@doi [\nat]
  {10.1038/328778a0}, \href
  {https://ui.adsabs.harvard.edu/abs/1987Natur.328..778P} {328, 778}

\bibitem[\protect\citeauthoryear{{Pushkarev} \& {Kovalev}}{{Pushkarev} \&
  {Kovalev}}{2012}]{pushkarev2012Tb}
{Pushkarev} A.~B.,  {Kovalev} Y.~Y.,  2012, \mn@doi [\aap]
  {10.1051/0004-6361/201219352}, \href
  {https://ui.adsabs.harvard.edu/abs/2012A&A...544A..34P} {544, A34}

\bibitem[\protect\citeauthoryear{{Pushkarev}, {Hovatta}, {Kovalev}, {Lister},
  {Lobanov}, {Savolainen}  \& {Zensus}}{{Pushkarev}
  et~al.}{2012}]{pushkarev2012}
{Pushkarev} A.~B.,  {Hovatta} T.,  {Kovalev} Y.~Y.,  {Lister} M.~L.,  {Lobanov}
  A.~P.,  {Savolainen} T.,   {Zensus} J.~A.,  2012, \mn@doi [\aap]
  {10.1051/0004-6361/201219173}, \href
  {https://ui.adsabs.harvard.edu/abs/2012A&A...545A.113P} {545, A113}

\bibitem[\protect\citeauthoryear{{Pushkarev}, {Kovalev}, {Lister}  \&
  {Savolainen}}{{Pushkarev} et~al.}{2017}]{pushkarev2017}
{Pushkarev} A.~B.,  {Kovalev} Y.~Y.,  {Lister} M.~L.,   {Savolainen} T.,  2017,
  \mn@doi [\mnras] {10.1093/mnras/stx854}, \href
  {https://ui.adsabs.harvard.edu/abs/2017MNRAS.468.4992P} {468, 4992}

\bibitem[\protect\citeauthoryear{{Ramakrishnan} et~al.,}{{Ramakrishnan}
  et~al.}{2016}]{ramakrishnan2016}
{Ramakrishnan} V.,  et~al., 2016, \mn@doi [\mnras] {10.1093/mnras/stv2653},
  \href {https://ui.adsabs.harvard.edu/abs/2016MNRAS.456..171R} {456, 171}

\bibitem[\protect\citeauthoryear{{Rani} et~al.,}{{Rani}
  et~al.}{2013}]{rani2013}
{Rani} B.,  et~al., 2013, \mn@doi [\aap] {10.1051/0004-6361/201321058}, \href
  {https://ui.adsabs.harvard.edu/abs/2013A&A...552A..11R} {552, A11}

\bibitem[\protect\citeauthoryear{{Readhead}}{{Readhead}}{1994}]{readhead1994}
{Readhead} A. C.~S.,  1994, \mn@doi [\apj] {10.1086/174038}, \href
  {https://ui.adsabs.harvard.edu/abs/1994ApJ...426...51R} {426, 51}

\bibitem[\protect\citeauthoryear{{Richards} et~al.,}{{Richards}
  et~al.}{2011}]{richards2011}
{Richards} J.~L.,  et~al., 2011, \mn@doi [\apjs] {10.1088/0067-0049/194/2/29},
  \href {https://ui.adsabs.harvard.edu/abs/2011ApJS..194...29R} {194, 29}

\bibitem[\protect\citeauthoryear{{Rybicki} \& {Lightman}}{{Rybicki} \&
  {Lightman}}{1979}]{rybicki1979}
{Rybicki} G.~B.,  {Lightman} A.~P.,  1979, {Radiative processes in
  astrophysics}

\bibitem[\protect\citeauthoryear{{Rybicki} \& {Lightman}}{{Rybicki} \&
  {Lightman}}{1986}]{rybicki1986}
{Rybicki} G.~B.,  {Lightman} A.~P.,  1986, {Radiative Processes in
  Astrophysics}

\bibitem[\protect\citeauthoryear{{Shepherd}}{{Shepherd}}{1997}]{shepherd1997}
{Shepherd} M.~C.,  1997, in {Hunt} G.,  {Payne} H.,  eds,  Astronomical Society
  of the Pacific Conference Series Vol. 125, Astronomical Data Analysis
  Software and Systems VI. p.~77

\bibitem[\protect\citeauthoryear{{Titarchuk}, {Seifina}, {Chekhtman}  \&
  {Ocampo}}{{Titarchuk} et~al.}{2020}]{titarchuk2020}
{Titarchuk} L.,  {Seifina} E.,  {Chekhtman} A.,   {Ocampo} I.,  2020, \mn@doi
  [\aap] {10.1051/0004-6361/201935576}, \href
  {https://ui.adsabs.harvard.edu/abs/2020A&A...633A..73T} {633, A73}

\bibitem[\protect\citeauthoryear{{T{\"u}rler} et~al.,}{{T{\"u}rler}
  et~al.}{1999}]{turler1999}
{T{\"u}rler} M.,  et~al., 1999, \mn@doi [\aaps] {10.1051/aas:1999125}, \href
  {https://ui.adsabs.harvard.edu/abs/1999A&AS..134...89T} {134, 89}

\bibitem[\protect\citeauthoryear{{Valtaoja}, {L{\"a}hteenm{\"a}ki},
  {Ter{\"a}sranta}  \& {Lainela}}{{Valtaoja} et~al.}{1999}]{valtaoja1999}
{Valtaoja} E.,  {L{\"a}hteenm{\"a}ki} A.,  {Ter{\"a}sranta} H.,   {Lainela} M.,
   1999, \mn@doi [\apjs] {10.1086/313170}, \href
  {https://ui.adsabs.harvard.edu/abs/1999ApJS..120...95V} {120, 95}

\bibitem[\protect\citeauthoryear{{Weaver} et~al.,}{{Weaver}
  et~al.}{2020}]{weaver2020}
{Weaver} Z.~R.,  et~al., 2020, \mn@doi [\apj] {10.3847/1538-4357/aba693}, \href
  {https://ui.adsabs.harvard.edu/abs/2020ApJ...900..137W} {900, 137}

\bibitem[\protect\citeauthoryear{{Weaver} et~al.,}{{Weaver}
  et~al.}{2022}]{weaver2022}
{Weaver} Z.~R.,  et~al., 2022, \mn@doi [\apjs] {10.3847/1538-4365/ac589c},
  \href {https://ui.adsabs.harvard.edu/abs/2022ApJS..260...12W} {260, 12}

\bibitem[\protect\citeauthoryear{{Wehrle} et~al.,}{{Wehrle}
  et~al.}{2012}]{wehrle2012}
{Wehrle} A.~E.,  et~al., 2012, \mn@doi [\apj] {10.1088/0004-637X/758/2/72},
  \href {https://ui.adsabs.harvard.edu/abs/2012ApJ...758...72W} {758, 72}

\makeatother
\end{thebibliography}

\appendix
\section{Jet Geometry} \label{sec:A:Jet_Geo}
Estimating the magnetic field strength of the jet using the SSA spectrum requires the angular size of the SSA emitting region (i.e., angular size at $\nu_{\rm m}$). The angular size of SSA regions, $d_{\rm m}$, can be interpolated (or extrapolated) from the model-fitted angular size of the radio core by using the jet geometry of the source. We reduced the same VLBA data that was used in \citet{kutkin2014} to obtain core sizes with their uncertainties. Uncertainties were estimated following \citet{fomalont1999}. Then the jet geometry ($W_{\rm core}$) of 3C~454.3 was modeled in logarithmic scale as follows:
    \begin{equation} \begin{aligned}
    \resizebox{.4 \textwidth}{!}{$
    {\rm log}(W_{\rm core})=\begin{cases} {\epsilon_{1}}\,{\rm log}(r(\nu)) + {\rm C}, & \nu \geq \nu_{\rm t} \\ {\epsilon_{2}}\,{\rm log}(r(\nu)) + (\epsilon_{1}-\epsilon_{2})\,{\rm log}(r_{\rm t}) + {\rm C}, & \nu < \nu_{\rm t}, \end{cases} \label{eq:A1:Jet_Geo} $}
    \end{aligned} \end{equation}
where $r_{\rm t}$ and $\nu_{\rm t}$ indicate the de-projected distance and frequency in units of pc and GHz at the jet geometry transition point, respectively. The parameter C is a constant. $\epsilon_{1}$ and $\epsilon_{2}$ are the geometrical indices ($W_{\rm core} \propto r(\nu)^{\epsilon_{1}}$ for $\nu \geq \nu_{\rm t}$ and $W_{\rm core} \propto r(\nu)^{\epsilon_{2}}$ for $\nu < \nu_{\rm t}$, and $\epsilon<1\colon \rm parabolic$ ; $\epsilon=1\colon \rm conical$ ; $\epsilon>1\colon \rm hyperbolic$).

In the fitting, the de-projected distance of a radio core at frequency $\nu$ is estimated by using the equation \citep{lobanov1998}$\colon r(\nu)={\frac{\Omega}{\rm sin \theta}} \nu^{-1/k}$, where $\Omega$ is the core position offset measurement between two frequencies $\nu_{1}$ and $\nu_{2}$, $k$ is the core-shift index, $\theta$ is the viewing angle. Following \citet{kutkin2014}, we adopt $\Omega=43\pm10~{\rm pc \, GHz}^{1/k}$, $k=0.7$ and $\theta=1.3^{\circ}$.

\begin{figure} \includegraphics[width=\columnwidth]{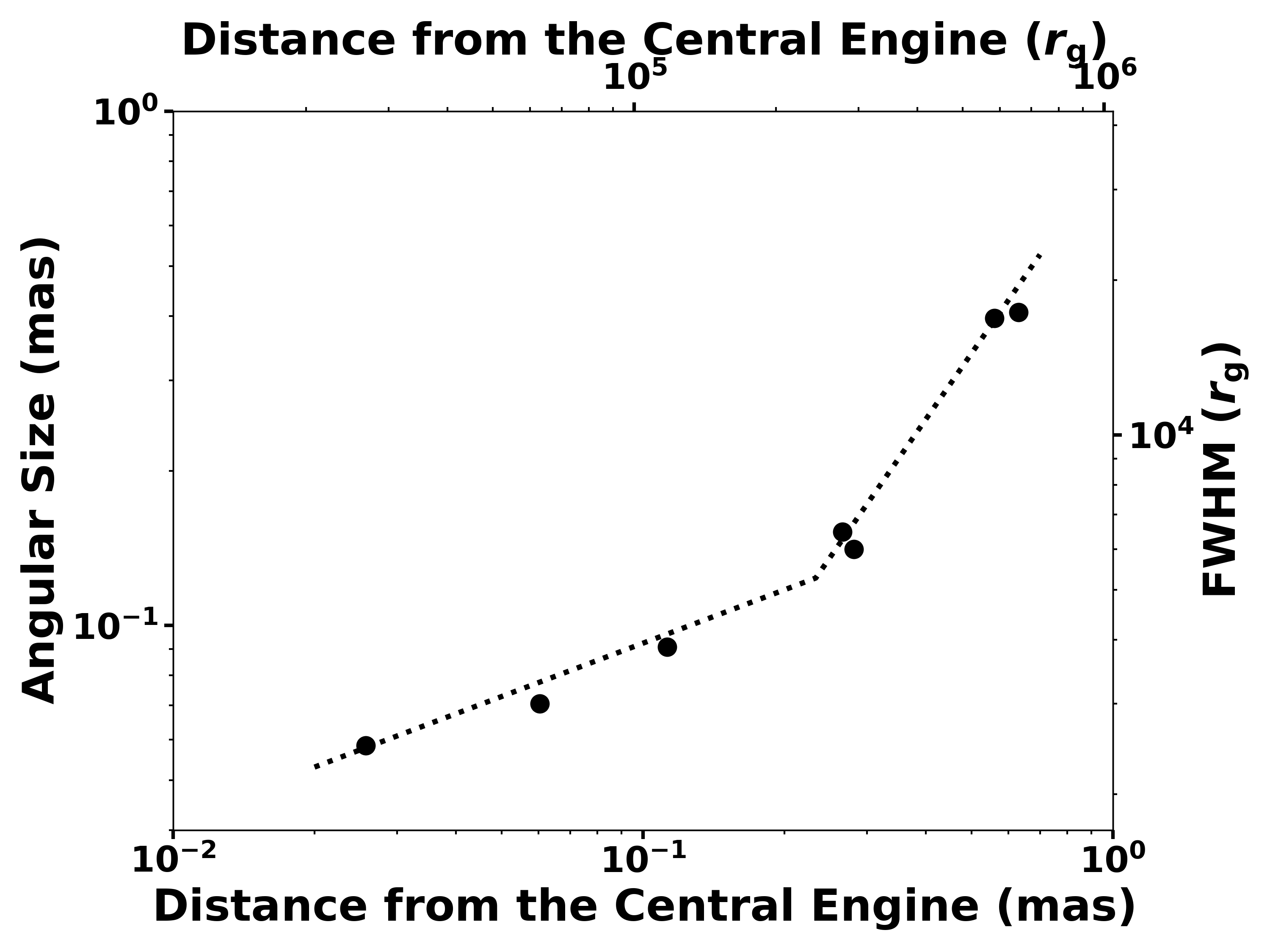}
    \caption{Estimated jet geometry using the VLBA core sizes from 43 to 4.6 GHz. The black dotted line indicates modeled jet geometry, and the black dots indicate measured core sizes from the October 2008 VLBA observation.} \label{fig:A1:Jet_Geo} \end{figure}

Figure \ref{fig:A1:Jet_Geo} shows the modeled jet geometry of 3C~454.3 with the measured core sizes. The fit parameters resulted in $\epsilon_{1}=0.35^{+0.09}_{-0.10}$, $\epsilon_{2}=1.32^{+0.30}_{-0.15}$, $r_{\rm t}=79^{+33}_{-15}~{\rm pc}$. \citet{kovalev2020} estimated a transition point of the jet geometry from nearby AGN sources, yielding $10^{5} - 10^{6}$ gravitational radii, $r_{\rm g}$ for a jet transition. Using the $M_{\rm SMBH}$ of 3C~454.3 assumed in this work, $r_{\rm t}$ is converted into $2.4 \times 10^5~r_{\rm g}$. This result is consistent with that estimated by \citet{kovalev2020}. Using the modeled jet geometry, the interpolated FWHM size of the SSA emitting region, $\theta_{\rm FWHM}$ is estimated and summarized in Table~\ref{table:3:SSA Parameters}.

% Don't change these lines
\bsp	% typesetting comment
\label{lastpage}
\end{document}